\documentclass[final,1p,times]{elsarticle}

\usepackage{lineno,hyperref}
\modulolinenumbers[5]

\usepackage{booktabs}
\usepackage{array,caption,threeparttable}
\usepackage[font=small,labelfont=bf]{caption}
\usepackage{amsmath,amssymb,amsfonts}
\usepackage{textcomp}

\def\BibTeX{{\rm B\kern-.05em{\sc i\kern-.025em b}\kern-.08em		T\kern-.1667em\lower.7ex\hbox{E}\kern-.125emX}}

\usepackage{graphicx}
\usepackage{epsfig}
\usepackage{subfigure}

\usepackage{array}

\usepackage{multirow}

\usepackage{float}
\usepackage{caption}
\usepackage{color}
\usepackage{verbatim}
\usepackage{bbding}
\usepackage{enumerate}
\usepackage{ntheorem}
\usepackage{cases}
\usepackage{booktabs} 
\usepackage{lscape}
\usepackage{makecell}

\usepackage{bm}

\usepackage{amsfonts}

\usepackage{mathrsfs}

\newtheorem{assumption}{Assumption}
\newtheorem{lemma}{Lemma}
\newtheorem{theorem}{Theorem}

\newtheorem{corollary}{Corollary}
\newtheorem{remark}{Remark}

\def\be{\begin{equation}}
\def\ee{\end{equation}}
\def\bea{\begin{eqnarray}}
\def\eea{\end{eqnarray}}
\newcommand{\re}[1]{(\ref{#1})}
\def\ne{ \nonumber \\ }
\def\nn{ \nonumber }
 \def\QED{~\rule[-1pt]{5pt}{5pt}\par\medskip}

\begin{document}

\begin{frontmatter}

\title{
Adaptive Neural Control with Desired
 Approximation:\\
An Integral Lyapunov Function Approach
}

\author{Mingxuan Sun
}

\author{Shengxiang Zou}
\ead{sxzouhz@hotmail.com}

\address{College of Information Engineering, Zhejiang University of
Technology, Hangzhou, 310023, China}

\begin{abstract}
The inherent approximation ability of neural networks plays an essential role in adaptive neural control,
where the prerequisite for existence of the compact set is crucial in the
control designs.
Instead of using practical system state,
in this paper,
the desired approximation approach is characterized to tackle such a problem,
where the desired state signal is required only as the input to the network.
An integral Lyapunov function-based
 adaptive controller is designed,
in the sense of the error tracking,
where
the treatment of the state-dependent input gain is adopted.
Theoretical results for the performance analysis
of the integral and incremental adaptation algorithms
are presented in details.
In particular,
the boundedness of the variables in the closed-loop is characterized,
while the transient performance of the output error is analytically quantified.
It is shown that the proposed control schemes assure that the tracking error converges to an adjustable set without any requirement on the knowledge of the region that the practical variables evolve,
and remove the requirement for the setting of initial conditions including system states and weight estimates.
\end{abstract}

\begin{keyword}
Approximation,
adaptive neural control,
desired compensation adaptive control,
error-tracking approach,
integral Lyapunov functions.
\end{keyword}

\end{frontmatter}


\section{Introduction}
\label{sec.introduction}

Typical learning mechanism,
featured by parametric estimation
for handling nonlinearities in system dynamics,
is involved in adaptive systems.
Adaptive control of nonlinear systems,
among others, has become an increasingly popular research area
and many significant advances have been made \cite{krstic95,ioannon06},
in developing state-/output-feedback control schemes,
through coping with linear-in-the-parameters uncertainties.
For most adaptive system designs,
the determination of regression models is generally dependent on the knowledge about the dynamics structure.
Both the high complexity and uncertainty involved in nonlinear dynamics may make the designs challenging.
Approximation tools,
such as neural networks (NNs) \cite{Ge01,Spooner02,Farrell06},
provide an efficient way to solve the problem,
which relax the requirement for the knowledge about the system dynamics,
and promote further developments in the context of adaptive control.

Representative works in the early stage of adaptive neural control were found in
\cite{Narendra90,Sanner92,Sadegh93}.
Among them,
attempts have been made
to cope with the possible controller singularity.
In \cite{Zhang00a},
an integral Lyapunov function was introduced for constructing the neuro-controller
so that such a problem is avoided.
The further improvement in the adaptive repetitive control design was presented, in \cite{sun06},
for a class of nonlinear parameterized systems.
For an adaptive neural control scheme,
the tracking performance is subject to the approximation error,
and the basic way for performance improvement is to choose sufficient enough neurons to reduce the error \cite{Ge01,Spooner02,Farrell06}.
The modification techniques are usually used,
in adaptive neural control designs,
to keep the weight estimates from growing without bound in the presence of system uncertainties and approximation errors,
and to enhance robustness of the closed-loop systems,
such as $\sigma$-modification \cite{ioannon06, Volyanskyy09},
projection operator,
and deadzone modification \cite{Ge01,Spooner02}.
The backstepping design techniques \cite{Ge01,WangW17}
have been widely applied to the nonlinear systems with triangular structures.
There were also attempts to extend to more general classes of triangular systems which have no affine appearance of variables to be used as control variables or virtual ones \cite{Ge02,Wang02}.
In \cite{ChenB16},
adaptive neural backstepping control strategies have been proposed for systems in nonstrict-feedback form,
in order to further relax the restrictions on the system structure.

The existing theoretical results may lead to a requirement for adaptive neural control:
The input variables of the NN undertaken should be kept in the compact set.
It is
a common recognition that
the compact set prerequisite plays an essential role for adaptive neural control.
Many reported works implicitly assumed that such a compact set always exists.
It would be a reasonable according to physical constraints, saturation, and restrictions imposed by performance and safety regulations.
The adaptive neural control is applicable,
when the system state within the region,
whereas beyond the region,
the NN fails to work,
and the adaptive bounding technique \cite{Farrell06} is helpful to assure that the state converges to the
region.
The constraint control methods for neural controller designs were reported in \cite{Tee09,Liu16,Song17},
in which Barrier Lyapunov functions
 provide an efficient way for assuring the compact set prerequisite.
More recently,
various adaptive NN-based nonlinear control schemes have been shown to efficiently tackle challenges such as
input saturation \cite{Ma15},
state constraints \cite{Tang16} and output constraints \cite{Song18}.

For coping with parametric uncertainties,
the desired compensation strategy,
using only the desired trajectory
quantities in the feedforward compensation,
was proposed, in \cite{sadegh90}, for trajectory tracking of robotic manipulators.
In \cite{yao09}, the desired
compensation strategy is applied in the framework of adaptive robust control,
for systems with linear-in-parameters uncertainties, in normal and strict-feedback forms.
where the
regressor and parameter estimation
depend on the reference trajectory only.
In this paper,
approximation-based control design is addressed by applying
the desired compensation strategy.
Here the desired compensation strategy lies twofold:
1) approximating the nonlinearity with neural networks, by applying only the
the desired trajectory quantities.
Throughout this paper, we call it the desired approximation (DA). And
2) the compensation for the nonlinearity provided by the DA
is applied in the control action.
The initial effort has been made and
the research result was reported in \cite{chenws08}, which
is closely related to our work presented in this paper.
In this paper, with the use of integral Lyapunov function proposed in \cite{sun18,sun23,zou23}, we provide an adaptive neural control design method, where both the integral and incremental adaptive algorithms are
suggested for parameter estimation.
The technical challenge regarding
the compact set prerequisite is addressed.

The main contributions of
this paper lie in three folds:

1) Taking into account that the control gain involved in the system undertaken is state-dependent,
an integral Lyapunov function is specifically constructed and used for the control designs in the sense of error tracking.

2) The DA-based adaptive neural control designs are conducted, together with the evaluation
of both transient and steady-state performance, for which a technical lemma is especially
given for characterizing and establishing the robust convergence performance of the closed-loop.

3) For the DA-based adaptive neural control, the incremental adaptive algorithms are exploited,
with the aid of which, the system uncertainties can be compensated and the boundedness
of parameter estimates can be guaranteed, as in the case of conventional integral adaptive methods.

\section{Preliminaries: Conventional Approximation Versus Desired Approximation}
\label{sect.da}
In this section,
we describe the desired approximation concept that is applicable to adaptive neural control designs.
Adaptive neural control underlines the principle of deterministic equivalence,
and utilizes the inherent approximation ability of neural networks.
It has been shown that,
by choosing sufficient neuros,
the used neural network can approximate smooth functions over a compact set to the desired accuracy.
There exist available approximation tools such as radial basis function networks,
multi-layer neural networks,
fuzzy systems,
wavelet networks,
etc.

By the neural approximation,
a smooth nonlinear function, $F: \Omega_x \subset R^n \rightarrow R$,
can be expressed as
\bea
    F(x) = {W^\ast}^T S(x) + \epsilon(x), & \forall x \in \Omega_x
    \label{F_NN}
\eea
where $W^\ast \in R^N$ is the ideal constant weight vector,
with $N$ being the number of neural network nodes,
and $S(x) \in R^N$ is the basis vector
consisting of a series of activation functions;
$\epsilon(x)$ is the approximation error
satisfying that
\[
\bar{\epsilon} > |\epsilon (x)|, \forall x \in \Omega_x
\]
and $\Omega_x$ is chosen to be a compact set
in which the state trajectories evolve.

We give the following observations and comments about the conventional approximation:

\begin{enumerate}[i)]
  \item The existing approximation theory
  has the prerequisite for existence of the compact set, which actually depends on the input data collected from the system undertaken.
  In practical applications,
  physical constraints and performance and safety specifications will limit the growth of the system state.
  There exists usually such a compact set, namely,
  $\Omega_x$,
  and the prerequisite can be met.
  However,
  from a theoretical point of view,
  many published works face the dilemma that
  establishing the boundedness of the sate variables of the closed-loop system under the bounded prerequisite.
  It is usually difficult to examine how the resultant tracking performance depends on the compact set,
  since it is only a pre-assumed one without giving its size.

  \item
  The tracking performance of many control schemes are determined by the upper bound on $\epsilon(x)$,
  while the value of $\epsilon(x)$ obviously depends on the number of neurons.
  An intuitive way for control performance improvement
  is to reduce the upper bound of $\epsilon(x)$.
  Theoretically,
  $\epsilon(x)$ relates to the set $\Omega_x$.
  As $\Omega_x$ is fixed,
  $\bar{\epsilon}$ decreases with the number of neurons,
  and the upper bound of the approximation error can be reduced by increasing neurons.
  To increase the number of neurons,
  however,
  $\bar{\epsilon}$ does not necessarily to decrease.
  As a matter of fact,
  most of the existing research results have not clearly given the range of the compact set,
  the range of the $\Omega_x$ may expand
  with the state trajectories evolving.
  The widening of the set region will result in a larger upper bound of the approximation error.

  \item The dimension of input signals used for NN training varies with practical situations,
  and it should be as less as possible,
  for reducing the burden of online computing.
  The conventional control designs make the input of NNs not only relate to the current measured state,
  but also contain more intermediate variables.
 Large input dimensions of NNs will bring a challenge to the setting of neuron number.
  A large number of weights will increase the computational burden for the network training.

  \item For the neural control strategies that take the system state as the input of NNs,
  $S(x)$ in \re{F_NN} depends on the actual state.
  As such,
  the measurement noise has an effect on the weight learning efficiency,
  leading to reduction of the approximation accuracy and the steady-state tracking performance.
\end{enumerate}

The above discussions motivate us to consider a desired approximation approach,
based on which the neural control design can be carried out,
without needing to consider the region that the practical variables evolve.

Let us decompose the nonlinear function $F(x)$
into two parts:
\[
F(x) = F_1(x,x_d) + F_2(x_d)
\]
with the difference $F_1(x,x_d) = F(x) - F(x_d)$ and the time-varying function $F_2(x_d) = F(x_d)$,
where $x_d$ is the given desired trajectory.
For tackling the difference $F(x) - F(x_d)$,
the norm-bound technique is appropriate,
by assuming
\bea
     |F(x) - F(x_d)| \le l_F(x,x_d)
    \label{F.DA1_1}
\eea
where $l_F(x,x_d)$ is a continuous function,
satisfying that $l_F(x,x_d)$ tends to zero,
as $x$ approaches to $x_d$.
The time-varying function $F_2(x_d) = F(x_d)$ only depends on the desired state $x_d$,
not related with the practical state $x$.
For dealing with it,
we apply the existing approximation tool as follows:
\bea
    F(x_d) = {W^\ast}^T S(x_d) + \epsilon(x_d)
    \label{F.DA1_2}
\eea
For the given desired trajectory $x_d$,
there exists a compact set, $\Omega_d \subset R^n$,
such that $x_d \in \Omega_d$.
According to the approximation property,
there exists an ideal weight vector $W^\ast \in R^N$ such that ${W^\ast}^T S(x_d)$ can approximate $F(x_d)$,
with the base vector $S(x_d)$
and
the approximation error $\epsilon(x_d)$,
bounded by $\bar{\epsilon}_d>0$.
The DA approach for tackling the unknown nonlinearity, $F(x)$,
can be expressed as follows:
\bea
    F(x) = F(x) - F(x_d) + {W^\ast}^T S(x_d) + \epsilon(x_d)
    \label{F.DA2}
\eea

We have the following additional comments on the desired approximation:

\begin{enumerate}[i)]
  \item The DA does not approximate the function $F(x)$ itself,
  but adopts the bounding technique (see \re{F.DA1_1}),
  and applies the neural approximation for the unknown function which only depends on the desired state vector (see \re{F.DA1_2}).
  The difficulty lies in how to tackle $F(x_d)$,
  which can be considered a time-varying parameter.
  The desired approximation offers a solution to its estimation problem.

  \item Compared with the conventional method \re{F_NN},
  DA avoids the approximation for $F(x)$,
  but approximates $F(x_d)$, as shown in \re{F.DA1_2}.
  That is,
  $x \in \Omega_x$ is not required and there is no need to determine the size of $\Omega_x$.
  One does not need to care about whether the state is kept within a compact set or not.
  As for \re{F.DA1_2},
  the range of $\Omega_d$ can be pre-specified and determined with ease,
  as it only depends on the desired state $x_d$.
  As such,
  the performance specifications can be realized by appropriate network training.

  \item It should be noted that the basis vector $S(x_d)$ is known,
  and we can calculate it off-line,
  in order to reducing the online computing burden.
  By using enough neurons,
  the approximation error $\epsilon(x_d)$ can be reduced dramatically,
  due to $\Omega_d$ being fixed.
  In addition,
  the DA can largely reduce the input dimension,
  and avoid the impact of measurement errors,
  since $x_d$,
  the input variable to the neural network,
  is given in advance.

\end{enumerate}

For further clarifying our motivations,
we refer to the desired compensation approach to control designs \cite{sadegh90,yao09}.
Let us consider the following system
\[
\dot x = F(x) + u
\]
where $F(x)$ represents the unknown nonlinearity;
and for the given desired trajectory $x_d$,
we have
\[
\dot x_d = F(x_d) + u_d
\]
where $u_d$ indicates the desired input.
Denoting by $e = x - x_d$ the tracking error,
the error dynamics can be expressed as
\[
\dot e = F(x) - F(x_d) - u_d +u
\]
It is observed that for control design,
there exist two kinds of uncertainties to be tackled:
the difference $F(x) - F(x_d)$ and the time-varying function $u_d$.
We use the norm-bound technique for dealing with the term $F(x) - F(x_d)$.
For handling $u_d$,
an iterative learning algorithm is able to be applicable,
when the tracking task is performed repetitively.
Here,
note that $u_d = F(x_d) - \dot x_d$.
We only need to handle $F(x_d)$, a function of  $x_d$,
due to the derivative of the desired state $x_d$ being computable,
for which we
apply an alternative approximation method in this paper.

The above discussion is for the input gain being one.
However,
the situation becomes difficulty,
when the input gain depends on the state.
To make the problem be solved completely,
in this paper,
we shall present a treatment by using an integral Lyapunov function,
originally given in \cite{sun18}.
The following lemma is helpful for forming an integral Lyapunov function.

\begin{lemma}
\label{Lemma_integral}
For the positive function $f(t)$, being continuously differentiable for all $t > 0$ and satisfying $f(0) = 0$, and the positive function
$g(t)$ is continuous for all $t \ge 0$, then
$
\int^t_0 g(f(s))f'(s)ds > 0
$ for all $t>0$.
\end{lemma}
{\bf Proof}
At first, we will appeal to the result for changing a variable in a Riemann integral.
Let us define
$F(t) = \int_0^t g(f(s)) f'(s) ds$
and
$G(t) = \int_{f(0)}^{t} g(s) ds $.
Then
$F'(t) = g(f(t)) f'(t)$.
Using the chain rule, the derivative of
$G(f(t) )$ with respect to $t$ equates to  $F'(t)$, i.e., $G'(f(t) ) =g(f(t)) f'(t)$.
Hence, $F(t)- G(f(t)) = C$, where $C$ is a constant.
Because of $F(0) = 0$ and $G(f(0)) =0$,
we obtain $C=0$ and $F(t) = G(f(t))$.

Then, it follows that by $f(0) = 0$,
\bea
G(f(t)) = \int_0^{f(t)} g(s) ds
\eea
Therefore,
$G(f(t)) > 0$, as $f(t) > 0 $ and $g(t) >0$ for all $t>0$,
and in turn $F(t) >0$ for all $t>0$.
\QED

In addition, the following lemma
\cite{sun23}
is
especially useful for the performance analysis to be presented.
\begin{lemma} Let $d_k$ be a sequence of real numbers. Suppose that for positive sequences $r_k$ and $s_k$,
\bea
r_k \le r_{k-1} - s_k + d_k
\label{kineq}
\eea
and $r_0$ is finite
and $s_k\rightarrow 0$ as
$r_k \rightarrow 0$.
Then, i)
$s_k$ is bounded for all $k$, and
$ \limsup \limits_{k \to \infty } s_k \le \bar d$,
 if $d_k$ satisfies that $|d_k| \le \bar d$, for all $k$; and
ii)
$ \lim \limits_{k \to \infty }
s_k = 0$,
if $ \lim \limits_{k \to \infty } d_k = 0$.
\label{keylem}
\end{lemma}

\section{Adaptive Neural Control Using Desired Approximation}
\label{sect.integral}
In this section,
the DA-based adaptive neural controller design is carried out,
where by using the error-tracking approach,
an integral Lyapunov function is adopted to cope with the state-dependent input gain.

Let us consider the class of SISO nonlinear systems described by
\bea
    \left\{
      \begin{array}{l}
        \dot{x}_i   = x_{i+1}, \quad i=1,2,\cdots,n-1 \\
        g(x)\dot{x}_{n} = f(x) + u \\
        y = x_1
      \end{array}
    \right.
    \label{sys}
\eea
where $x=[x_1,x_2,\cdots,x_n]^T \in R^n$ is the state vector,
$u \in R$ and $y \in R$ are the scalar input and output of the system,
respectively.
Both $f(\cdot)$ and $g(\cdot)$ are continuously differentiable nonlinear functions.

For a given reference signal,
$y_d(t)$,
with $y_d(t),\dot{y}_d(t),\cdots$, and $y^{(n)}_d(t)$ being bounded,
let us denote by $e = x - x_d = [e_1, e_2, \cdots, e_n]^T$ the state error,
where $x_d = [y_d, \dot{y}_d, \cdots,$ $y^{(n-1)}_d]^T $ is the desired state.
The control objective is to find the control input such that the practical output tracks the given reference signal as closely as possible,
while all the variables in the closed-loop system are bounded.
To tackle the objective,
let us introduce the filtered error,
$e_f$,
defined as $e_f = [\Lambda^T \quad 1]e$,
where $\Lambda = [\lambda_1, \lambda_2, \cdots, \lambda_{n-1}]^T$ is chosen such that polynomial $s^{n-1} + \lambda_{n-1} s^{n-2} + \cdots + \lambda_1$ is Hurwitz.
According to the definition of $e_f$,
it also defines a proper stable transfer function,
$H(s) = (s^{n-1} + \lambda_{n-1} s^{n-2} + \cdots + \lambda_2 s + \lambda_1)^{-1}$, such that $e_1 = H(s) e_f$.
Obviously,
$e$ will converge to zero,
as $e_f$ approaches zero.
At the same time,
the output error ($e_1 = y - y_d$) converges to zero too.
The $e_f$-dynamics can be given as,
\bea
    g(x) \dot{e}_f = f(x) + g(x) \left([0 ~~ \Lambda^T]e - y^{(n)}_d\right) + u
    \label{dot.ef}
\eea

\begin{assumption}
\label{assum1}
The sign of function $g(x)$ is known.
\end{assumption}

\begin{remark}
Assumption \ref{assum1} implies that the control gain,
$g(x)$,
is strictly either positive or negative,
and $g(x)$ does not pass through the origin,
indicating system \re{sys} being controllable.
Without lose of generality,
throughout of this paper,
it is assumed that $g(x) > 0$.
The following analysis is easy to modify for system \re{sys} with $g(x) < 0$.
\end{remark}

\begin{assumption}
\label{assum2}
There exists a known nonnegative function $l_h(\chi_1, \chi_2)$ such that $h(\cdot)$ satisfies $|h(\chi_1) - h(\chi_2)| \le l_h(\chi_1,\chi_2)$,
for $\chi_1,\chi_2 \in R^n$,
and $l_h(\chi_1,\chi_2)$ tends to zero,
as $\chi_1$ tends to $\chi_2$,
for $h \in \{f,g\}$.
\end{assumption}

\subsection{Desired error trajectories}

We begin with the description for the error-tracking approach that is used for the neural controller design to be presented.
Let us denote by $\tilde{e}_f$ the error between $e_f$ and $e^\ast_f$,
where $e^\ast_f$ indicates the desired filtered error trajectory.
It follows from \re{dot.ef} that
\bea
g(x) \dot{\tilde{e}}_f = f(x) + g(x) \nu + u
\label{tilde_e1}
\eea
with $\nu = [0 ~~ \Lambda^T]e - y^{(n)}_d - \dot{e}^\ast_f$.

Now the control objective can be restated as follows:
find the control input such that $e_f(t)$ follows $e^\ast_f(t)$ on $[0,+\infty)$.
The transient and steady-state specifications for the filtered error are the requirements for $e^\ast_f$.
To achieve the error-tracking objective,
the following assumption is made,
a restriction on the setting of initial value of the desired filtered error trajectory, $e^\ast_f(t)$.

\begin{assumption}
\label{assum3}
The initial value, $e^\ast_f(0)$, is set to satisfy $e^\ast_f(0) = e_f(0)$.
\end{assumption}

\begin{remark}
Assumption \ref{assum3} implies that
\bea
\tilde{e}_f(0) = 0
\label{eq.ef0}
\eea
which is only a restriction on the initial value of the desired filtered error trajectory,
but not pose any requirement for the filtered error itself.
What's more,
the initial value of the state error is allowed to be given arbitrarily but bounded.
With intuitive and clear requirements for the attenuation behavior,
one can form the desired filtered error trajectory $e^\ast_f$ with ease.
To satisfy Assumption \ref{assum3},
$e^\ast_f(t)$ can be chosen in the form of
\bea
e^\ast_f(t) = e_f(0) \zeta(t)
\label{eq.efast}
\eea
Here $\zeta(t)$ is taken as a smooth and monotonically decreasing function on $[0,\Delta]$,
and satisfies
$\zeta(0) = 1$ and $\zeta(t) = 0$ for $t \in [\Delta,+\infty)$,
where $\Delta$ is the setting moment to connect the beginning position and the desired trajectory.
It is seen that $e^\ast_f(t)$ is continuously differentiable.
The initial filtered error $e_f(0)$ is determined by the actual state trajectory,
while $\zeta(t)$ and $\Delta$ can be set to assure the convergence performance of the desired filtered error trajectory.
Once $e_f(t)$ converges to the pre-specified error trajectory $e^\ast_f(t)$,
the output $y(t)$ will track $y_d(t)$ as $t \rightarrow \infty$.
\end{remark}

\subsection{DA-based adaptive neural control with integral adaptation}

For the control design,
the $\tilde{e}_f$-dynamics \re{tilde_e1} is written as
\begin{align}
    g(x)\dot{\tilde{e}}_f
  = &f(x) - f(x_d) + {W^\ast_f}^T S_f(x_d) + \epsilon_f(x_d) + u\ne
   &+\nu\left(g(x) - g(x_d) + {W^\ast_{g}}^T S_{g}(x_d) + \epsilon_{g}(x_d)\right)
  \label{tilde_e2}
\end{align}
with the approximation errors $\epsilon_f(x_d)$ and $\epsilon_{g}(x_d)$ bounded by
$\bar{\epsilon}_f>0$ and $\bar{\epsilon}_{g}>0$,
respectively.
For the error system \re{tilde_e2},
we propose the following adaptive neural control law:
\bea
    u = -\tilde{e}_f\left(\frac{\varrho^2}
      {\sqrt{\tilde{e}^2_f \varrho^2 + \varepsilon^2_\varrho}} + \kappa\right)
    - \hat{W}^T_f S_f
    - \hat{W}^T_{g} S_{g} \nu
    \label{u_DA}
\eea
where
$\varrho = l_f(x,x_d) + l_{g}(x,x_d) |\nu| + |\tilde{e}_f|(1 + \nu^2)/4\eta$,
and $\kappa, \varepsilon_\varrho, \eta > 0$ are adjustable,
and $\hat{W}_f$ and $\hat{W}_{g}$ are the estimates for $W^\ast_f$ and $W^\ast_{g}$, respectively.
The adaptation laws for updating $\hat{W}_f$ and $\hat{W}_{g}$ are as follows:
\begin{align}
\dot{\hat{W}}_f =& \Gamma_f \left(S_f(x_d)\tilde{e}_f - \sigma_f \hat{W}_f\right)
    \label{hatW_f_int}\\
\dot{\hat{W}}_{g}=&\Gamma_{g}\left(S_{g}(x_d)\tilde{e}_f\nu-\sigma_{g}\hat{W}_{g}\right)
    \label{hatW_g_int}
\end{align}
where
$\Gamma_f$ and $\Gamma_{g}$ are positive definite diagonal matrices and
$\sigma_f$ and $\sigma_{g}$ are positive constants.

With the control law \re{u_DA},
the $\tilde{e}_f$-dynamics \re{tilde_e3} becomes
\begin{align}
    g(x)\dot{\tilde{e}}_f
  =&f(x) - f(x_d) - {\tilde{W}_f}^T S_f(x_d) + \epsilon_f(x_d)+\nu\left(g(x)-g(x_d) - {\tilde{W}_{g}}^TS_{g}(x_d)+ \epsilon_{g}(x_d)\right)\ne
   &-\tilde{e}_f\left(\frac{\varrho^2}
   {\sqrt{\tilde{e}^2_f \varrho^2 + \varepsilon^2_\varrho}} + \kappa\right)
  \label{tilde_e3}
\end{align}
 with $\tilde{W}_h = \hat{W}_h - W^\ast_h, h \in \{f,g\}$, being the estimation errors.

The tracking performance of the closed-loop system is summarized in the following theorem.
\begin{theorem}
\label{thm.1}
Consider the adaptive system described by system \re{sys},
 neural controller \re{u_DA} and  weight adaptation laws \re{hatW_f_int}--\re{hatW_g_int},
under Assumptions \ref{assum1}--\ref{assum3}.
Then for any bounded initial conditions,
all the signals in the closed-loop system are bounded,
and the state $x$ remains in
$\Omega_x = \{x(t) \big| |e_f(t)| \le |e_f(0)| +B_{e_f}, x_d \in \Omega_d \}$,
$t \ge T$,
with $T$ to be specified and
$B_{e_f} = \kappa^{-1/2}
  \big(\varepsilon_\varrho
+ (\sigma_f \|W^\ast_f\|^2 + \sigma_{g} \|W^\ast_{g}\|^2 )/ 2
+ \eta \bar{\epsilon}^2_f
+ \eta \bar{\epsilon}^2_{g}\big)^{1/2}$.
Moreover,
$x(t) \in \bar{\Omega}_x = \{x(t) \big| |e_f(t)| \le B_{e_f}, x_d \in \Omega_d \}$,
as $t \ge T_1 = \max\{T,\Delta\}$.
The mean-square output error satisfies
$\frac{1}{t} \int^t_0 e^2_1(s) ds \le \frac{1}{t} (c_1 \varpi_0 + c_2) + 2 c_1 B^2_{e_f}$,
$t > 0$,
with
computable constants $c_1, c_2 > 0$, and
$\varpi_0 = \kappa^{-1}(\tilde{W}^T_f(0) \Gamma^{-1}_f \tilde{W}_f(0) + \tilde{W}^T_{g}(0) \Gamma^{-1}_{g} \tilde{W}_{g}(0)) + 2 e^2_f(0) \Delta$.
\end{theorem}
{\bf Proof}
Let us consider the integral Lyapunov function candidate,
$
L(t)= V(t)
    +\frac{1}{2} \tilde{W}^T_f \Gamma^{-1}_f \tilde{W}_f
    +\frac{1}{2} \tilde{W}^T_{g} \Gamma^{-1}_{g} \tilde{W}_{g}
$
with $V(t) = \int^t_0 g(x(s))\frac{d}{ds}\left(\frac{1}{2}\tilde{e}^2_f(s)\right)ds$.
In view of $\tilde{e}_f(0) = 0$,
the positivity of $V$ can be established,
with the aid of Lemma \ref{Lemma_integral}.
The derivative of $L$ with respect to time along \re{tilde_e3} is calculated as
\begin{align}
  \dot{L}
 =& \tilde{e}_f
    \Bigg(f(x) - f(x_d) - {\tilde{W}_f}^T S_f(x_d) + \epsilon_f(x_d)\ne
   &
    +\nu\left(g(x)-g(x_d) - {\tilde{W}_{g}}^TS_{g}(x_d)+ \epsilon_{g}(x_d)\right)
    -\kappa  \tilde{e}_f\ne
   &
    - \frac{\tilde{e}_f \varrho^2}{\sqrt{\tilde{e}^2_f \varrho^2 + \varepsilon^2_\varrho}}
    \Bigg)
   + \tilde{W}^T_f \Gamma^{-1}_f \dot{\hat{W}}_f
   + \tilde{W}^T_{g} \Gamma^{-1}_{g} \dot{\hat{W}}_{g}\ne
\le&
    |\tilde{e}_f|
    \Bigg(l_f(x,x_d)
        + l_{g}(x,x_d) |\nu|
        + \bar{\epsilon}_f
        + \bar{\epsilon}_{g} |\nu|
    \Bigg)\ne
   &
   - \kappa \tilde{e}^2_f
   - \frac{\tilde{e}^2_f \varrho^2}{\sqrt{\tilde{e}^2_f \varrho^2 + \varepsilon^2_\varrho}}
   + \tilde{W}^T_f \left(\Gamma^{-1}_f \dot{\hat{W}}_f - S_f(x_d) \tilde{e}_f\right)\ne
   &
   + \tilde{W}^T_{g}
     \left(\Gamma^{-1}_{g} \dot{\hat{W}}_{g} - S_{g}(x_d) \tilde{e}_f \nu\right)\nn
\end{align}

By virtue of Young's inequality,
there is a constant $\eta > 0$ such that
\begin{align}
    &- \frac{\tilde{e}^2_f}{4 \eta}
    + \bar{\epsilon}_f |\tilde{e}_f|
\le
      \eta \bar{\epsilon}^2_f\ne
    &- \frac{\tilde{e}^2_f \nu^2}{4 \eta}
    + \bar{\epsilon}_{g} |\tilde{e}_f| |\nu|
\le
      \eta \bar{\epsilon}^2_{g}\nn
\end{align}
It follows from the definition of $\varrho$ in \re{u_DA} that
\begin{align}
\dot{L}
\le&
   - \frac{1 + \nu^2}{4 \eta} \tilde{e}^2_f
   + |\tilde{e}_f| \varrho
   + |\tilde{e}_f| \bar{\epsilon}_f
   + |\tilde{e}_f| \bar{\epsilon}_{g} |\nu|\ne
   &
   - \kappa \tilde{e}^2_f
   - \frac{\tilde{e}^2_f \varrho^2}{\sqrt{\tilde{e}^2_f \varrho^2 + \varepsilon^2_\varrho}}
   + \tilde{W}^T_f \left(\Gamma^{-1}_f \dot{\hat{W}}_f - S_f(x_d) \tilde{e}_f\right)\ne
   &
   + \tilde{W}^T_{g}
     \left(\Gamma^{-1}_{g} \dot{\hat{W}}_{g} - S_{g}(x_d) \tilde{e}_f \nu\right)\ne
\le&
   - \kappa \tilde{e}^2_f
   + |\tilde{e}_f| \varrho
   - \frac{\tilde{e}^2_f \varrho^2}{\sqrt{\tilde{e}^2_f \varrho^2 + \varepsilon^2_\varrho}}
   + \eta \bar{\epsilon}^2_f
   + \eta \bar{\epsilon}^2_{g}\ne
   &
   + \tilde{W}^T_f \left(\Gamma^{-1}_f \dot{\hat{W}}_f - S_f(x_d) \tilde{e}_f\right)\ne
   &
   + \tilde{W}^T_{g}
     \left(\Gamma^{-1}_{g} \dot{\hat{W}}_{g} - S_{g}(x_d) \tilde{e}_f \nu\right)\nn
\end{align}
To proceed, we note that
\begin{align}
- \frac{\tilde{e}^2_f \varrho^2}{\sqrt{\tilde{e}^2_f \varrho^2 + \varepsilon^2_\varrho}}
+ |\tilde{e}_f| \varrho
=&
 |\tilde{e}_f| \varrho
 \frac{\sqrt{\tilde{e}^2_f \varrho^2 + \varepsilon^2_\varrho} - |\tilde{e}_f| \varrho}
      {\sqrt{\tilde{e}^2_f \varrho^2 + \varepsilon^2_\varrho}}\ne
\le&
 \sqrt{\tilde{e}^2_f \varrho^2 + \varepsilon^2_\varrho} - |\tilde{e}_f| \varrho\ne
\le&
 \sqrt{\tilde{e}^2_f \varrho^2} + \sqrt{\varepsilon^2_\varrho} - |\tilde{e}_f| \varrho
= \varepsilon_\varrho \nn
\end{align}
and applying the weight adaptation laws \re{hatW_f_int}--\re{hatW_g_int} leads to
\begin{align}
\dot{L}
\le&
   - \kappa \tilde{e}^2_f
   - \frac{\tilde{e}^2_f \varrho^2}{\sqrt{\tilde{e}^2_f \varrho^2 + \varepsilon^2_\varrho}}
   + |\tilde{e}_f| \varrho
   + \eta \bar{\epsilon}^2_f
   + \eta \bar{\epsilon}^2_{g}\ne
   &
   - \sigma_f \tilde{W}^T_f \hat{W}_f
   - \sigma_{g} \tilde{W}^T_{g} \hat{W}_{g}
   - \kappa \tilde{e}^2_f
   + \varepsilon_\varrho
   + \eta \bar{\epsilon}^2_f
   + \eta \bar{\epsilon}^2_{g}
   - \sigma_f \tilde{W}^T_f \hat{W}_f
   - \sigma_{g} \tilde{W}^T_{g} \hat{W}_{g}\nn
\end{align}

Using the fact that
\begin{align}
   - \sigma_h \tilde{W}^T_h \hat{W}_h
=& - \sigma_h \tilde{W}^T_h \tilde{W}_h - \sigma_h \tilde{W}^T_h W^\ast_h\ne
\le&
   - \frac{\sigma_h}{2} \|\tilde{W}_h\|^2
   + \frac{\sigma_h}{2} \|W^\ast_h\|^2, \quad h \in \{f,g\}\nn
\end{align}
gives rise to
\begin{align}
\dot{L}
\le&
   - \kappa \tilde{e}^2_f
   - \frac{\sigma_f}{2} \|\tilde{W}_f\|^2
   - \frac{\sigma_{g}}{2} \|\tilde{W}_{g}\|^2\ne
   &
   + \varepsilon_\varrho
   + \eta \bar{\epsilon}^2_f
   + \eta \bar{\epsilon}^2_{g}
   + \frac{\sigma_f}{2} \|W^\ast_f\|^2
   + \frac{\sigma_{g}}{2} \|W^\ast_{g}\|^2
\label{dotL.key}
\end{align}

Define
$ \Omega_{\tilde{e}_f}
= \{\tilde{e}_f(t) \big|
          |\tilde{e}_f(t)| \le \kappa^{-1/2}
 (\varepsilon_\varrho
+ \eta \bar{\epsilon}^2_f
+ \eta \bar{\epsilon}^2_{g}
+ \sigma_f \|W^\ast_f\|^2 / 2
+ \sigma_{g} \|W^\ast_{g}\|^2 / 2)^{1/2} = B_{e_f}\}$.
Then, the set $\Omega_{\tilde{e}_f}$ is a compact set,
and
$\dot{L}$ is negative outside this set.
As such,
$|\tilde{e}_f|$ will decrease and finally converge within the bound.
This establishes the UUB property of $\tilde{e}_f$,
implying that there exists the time constant $T$ such that $\tilde{e}_f(t) \in \Omega_{\tilde{e}_f}$ for $t \ge T$.
Similarly, from \re{dotL.key}, the UUB property
of both $\hat{W}_f$ and $\hat{W}_{g}$ can also be guaranteed.
For bounded initial conditions,
the boundedness of $e_f(t)$ can be immediately established by
$|e_f(t)| \le |e^\ast_f(t)| + |\tilde{e}_f(t)| \le |e_f(0)| + B_{e_f}, \forall t \ge T$.
Since $e^\ast_f(t) = 0, \forall t \ge \Delta$,
then $|e_f(t)| \le B_{e_f}$ as $t \ge T_1$.
In turn,
$x(t)$ remains in $\Omega_{x}$,
for $t \in [T,+\infty)$,
and $x(t) \in \bar{\Omega}_{x}$ as $t \ge T_1$.
Due to boundedness of $e_f, x_d, y^{(n)}_d, \hat{W}_f$ and $\hat{W}_{g}$,
it is easy from \re{u_DA} to establish the boundedness of the control input, $u$.

Now,
integrating \re{dotL.key} over $[0,t]$ results in
\begin{align}
   \int^t_0 \tilde{e}^2_f(s) ds
\le&\frac{1}{\kappa}\left(L(0) - L(t)\right)
+ \kappa^{-1}
  \Big(\varepsilon_\varrho
+ \eta \bar{\epsilon}^2_f
+ \eta \bar{\epsilon}^2_{g}\ne
&
+ \frac{\sigma_f}{2} \|W^\ast_f\|^2
+ \frac{\sigma_{g}}{2} \|W^\ast_{g}\|^2\Big)t\nn
\end{align}
Note that $L(0)$ is only related to the initial values of the weight estimation errors.
According to the non-negativity of $L$,
\begin{align}
    &\int^t_0 \tilde{e}^2_f(s) ds
\le\frac{1}{2\kappa} \tilde{W}^T_f(0) \Gamma^{-1}_f \tilde{W}_f(0)
  + \frac{1}{2\kappa} \tilde{W}^T_{g}(0) \Gamma^{-1}_{g} \tilde{W}_{g}(0)\ne
&\qquad
  + \kappa^{-1}
  \Big(\varepsilon_\varrho
+ \eta \bar{\epsilon}^2_f
+ \eta \bar{\epsilon}^2_{g}
+ \frac{\sigma_f}{2} \|W^\ast_f\|^2
+ \frac{\sigma_{g}}{2} \|W^\ast_{g}\|^2\Big)t
\nn
\end{align}
In addition, utilizing that $|e^\ast_f(t)| \le |e_f(0)|$,
for $t \ge 0$,
and $e^\ast_f(t) = 0$,
as $t \ge \Delta$, we obtain
\begin{align}
    \int^t_0 e^2_f(s) ds
\le&2 \int^t_0 \tilde{e}^2_f(s) ds + 2 \int^\Delta_0 {e^\ast_f}^2(s) ds\ne
\le&\frac{1}{\kappa} \tilde{W}^T_f(0) \Gamma^{-1}_f \tilde{W}_f(0)
+ \frac{1}{\kappa} \tilde{W}^T_{g}(0) \Gamma^{-1}_{g} \tilde{W}_{g}(0)\ne
&+ 2 e^2_f(0) \Delta
 + 2 \kappa^{-1}
  \Big(\varepsilon_\varrho
+ \eta \bar{\epsilon}^2_f
+ \eta \bar{\epsilon}^2_{g}
+ \frac{\sigma_f}{2} \|W^\ast_f\|^2
+ \frac{\sigma_{g}}{2} \|W^\ast_{g}\|^2\Big)t
\nn
\end{align}
According to the relationship between $e_1$ and $e_f$, there exist positive constants $c_1$ and
$c_2$ such that
 $\int^t_0 e^2_1(s) ds
\le c_1 \int^t_0 e_f^2(s) ds + c_2$, under which
the mean-square output error can be established by observing that
\begin{align}
    \int^t_0 e^2_1(s) ds
\le& c_1\Bigg(\frac{1}{\kappa} \tilde{W}^T_f(0) \Gamma^{-1}_f \tilde{W}_f(0)
  + \frac{1}{\kappa} \tilde{W}^T_{g}(0) \Gamma^{-1}_{g} \tilde{W}_{g}(0)\ne
&
  + 2 e^2_f(0) \Delta
  + 2 \kappa^{-1}
  \Big(\varepsilon_\varrho
+ \eta \bar{\epsilon}^2_f
+ \eta \bar{\epsilon}^2_{g}
+ \frac{\sigma_f}{2} \|W^\ast_f\|^2
+ \frac{\sigma_{g}}{2} \|W^\ast_{g}\|^2\Big)t\Bigg) + c_2
    \label{e_ms1}
\end{align}
This completes the proof.
\QED

\begin{remark}
The mean-square error given by Theorem \ref{thm.1} reveals the transient response of the closed-loop system,
which reflects the average tracking performance over the whole control process.
Furthermore,
taking the limit leads to $\lim_{t \rightarrow \infty} \frac{1}{t} \int^t_0 e^2_1(s) ds \le 2 c_1 B^2_{e_f}$,
implying that
the output error $e_1(t)$ will converge to the neighborhood of the origin in the sense of mean-square,
and the radius of the neighborhood is adjustable.
\end{remark}

\begin{corollary}
\label{cor_filtered_error}
Let the controller \re{u_DA} together with the adaptation laws \re{hatW_f_int}--\re{hatW_g_int}
be applied to system \re{sys},
under Assumptions \ref{assum1}--\ref{assum3}.
Then,
the filtered error satisfies that
$\lim_{t \rightarrow \infty} \sup \int^t_{t-\tau} e^2_f (s) ds \le \tau B^2_{e_f}$,
where $\tau > 0$ is a constant and $B_{e_f}$ is defined in Theorem \ref{thm.1}.
\end{corollary}
{\bf Proof}
As in the proof of Theorem \ref{thm.1},
again integrating \re{dotL.key} over $[t-\tau, t]$ gives rise to
\begin{align}
   L(t) - L(t-\tau)
\le&
- \kappa \int^t_{t-\tau} \tilde{e}^2_f(s) ds
+ \Big(\varepsilon_\varrho
   + \eta \bar{\epsilon}^2_f
   + \eta \bar{\epsilon}^2_{g}
   + \frac{\sigma_f}{2} \|W^\ast_f\|^2
   + \frac{\sigma_{g}}{2} \|W^\ast_{g}\|^2\Big) \tau\nn
\end{align}
For the case where $L(t_0), t_0 \in [0, \tau]$ is finite,
in the light of Lemma \ref{keylem},
the integral term $\int^t_{t-\tau} \tilde{e}^2_f (s) ds$ could enter the specified bound $\kappa^{-1} \tau(\varepsilon_\varrho
+ \eta \bar{\epsilon}^2_f
+ \eta \bar{\epsilon}^2_{g}
+ \sigma_f \|W^\ast_f\|^2 / 2
+ \sigma_{g} \|W^\ast_{g}\|^2 / 2)$,
as $t \rightarrow +\infty$.
Since $\lim_{t \rightarrow \infty} e^\ast_f(t) = 0$,
\begin{align}
\lim_{t \rightarrow \infty} \sup \int^t_{t-\tau} e^2_f (s) ds
\le&
  \frac{\tau}{\kappa}
  \Big(\varepsilon_\varrho
   + \eta \bar{\epsilon}^2_f
   + \eta \bar{\epsilon}^2_{g}
   + \frac{\sigma_f}{2} \|W^\ast_f\|^2
   + \frac{\sigma_{g}}{2} \|W^\ast_{g}\|^2\Big)
  \nn
\end{align}
In order to obtain the above convergence performance of the filtered error,
we have to consider the finiteness of $L(t_0)$.
It follows from \re{dotL.key} that
\begin{align}
\dot{L}(t_0)
\le \varepsilon_\varrho
   + \eta \bar{\epsilon}^2_f
   + \eta \bar{\epsilon}^2_{g}
   + \frac{\sigma_f}{2} \|W^\ast_f\|^2
   + \frac{\sigma_{g}}{2} \|W^\ast_{g}\|^2 \nn
\end{align}
Since $W^\ast_f$ and $W^\ast_{g}$ are bounded,
$L(t_0)$ is bounded on the interval $[0,\tau]$.
By invoking Lemma \ref{keylem},
the convergence result of the filtered error can be established.
This completes the proof.
\QED

Corollary \ref{cor_filtered_error}
establishes the asymptotic convergence of the filtered error in the sense of $L_2$ over $[t-\tau, t]$, as $t \rightarrow \infty$, and in turn assures the asymptotic convergence of the tracking error,
where Lemma \ref{keylem} plays a key role to finalize the analysis.

\begin{remark}
In the proof for Theorem \ref{thm.1}, we adopt the integral Lyapunov function, $L(t)$,
for the case where the $g(x)$ in \re{sys} depends on the state $x$.
This significantly simplifies and shortens
the control design and analysis.
It has been shown that such a Lyapunov function can also effectively solve the so-called control singularity problem that is usually encountered in the conventional designs \cite{sun18}.
\end{remark}

\section{DA-based Adaptive Neural Control with Incremental Adaptation}

Instead of using the conventional integral adaptation, in this section,
the following incremental adaptive mechanisms are proposed and employed for updating the weight estimates, $\hat{W}_f$ and $\hat{W}_g$,
\begin{align}
   (1 + \sigma_f \Gamma_f) \hat{W}_f(t)
=& \hat{W}_f(t - \tau) + \Gamma_f \tilde{e}_f(t) S_f(x_d)
\label{hatW_fDA} \\
   (1 + \sigma_g \Gamma_g) \hat{W}_g(t)
=& \hat{W}_g(t - \tau) + \Gamma_g \tilde{e}_f(t) \nu(t) S_g(x_d)
\label{hatW_gDA}
\end{align}
where $\Gamma_f$ and $\Gamma_g$ are positive definite diagonal matrices,
$\sigma_f$ and $\sigma_g$ are positive constants,
and $\tau > 0$ is the learning period.
Note that with the control law \re{u_DA} together with
\re{hatW_fDA}-\re{hatW_gDA},
the $\tilde{e}_f$-dynamics is in the same form
as \re{tilde_e3}.

\begin{remark}
A major obstacle for the implementation is
to accurately realize
integral adaptive laws \re{hatW_f_int} and \re{hatW_g_int}.
To avoid the numerical integration for implementation,
incremental adaption \re{hatW_fDA} and \re{hatW_gDA} are suggested for parameter adaptation.
As a substitute for the integral adaptive laws \re{hatW_f_int} and \re{hatW_g_int},
in what follows,
\re{hatW_fDA} and \re{hatW_gDA} are demonstrated to work well.
In particular,
typical setting in the implementation
is to take $\tau$ as the sampling period or an integer multiple of the sampling period.

\end{remark}

The major result about the tracking performance of the closed-loop system is summarized in the following theorem.
\begin{theorem}
\label{thm.2}
Consider the adaptive system described by system \re{sys},
neural controller \re{u_DA} and  weight adaptation laws \re{hatW_fDA} and \re{hatW_gDA},
under Assumptions \ref{assum1}--\ref{assum2}.
Then
\begin{enumerate}
\item
all the signals in the closed-loop system are bounded;

\item
the state $x$ remains in
$\Omega_x = \Big\{x(t) \big| |e_f(t)| \le |e_f(0)| + \kappa^{-1/2}$ $\epsilon(\epsilon_\varrho, \bar{\epsilon}_f, \bar{\epsilon}_g, \sigma_f, \sigma_g)^{1/2}, x_d \in \Omega_d\Big\}$,
$t \ge T$,
with $T$ to be specified and
$\epsilon(\epsilon_\varrho, \bar{\epsilon}_f, \bar{\epsilon}_g, \sigma_f, \sigma_g) = \epsilon_\varrho
+ \eta \bar{\epsilon}^2_f
+ \eta \bar{\epsilon}^2_g
+ (\sigma_f \|W^\ast_f\|^2 + \sigma_g \|W^\ast_g\|^2)/ 2$.
Moreover,
$x(t) \in \bar{\Omega}_x = \Big\{x(t) \big| |e_f(t)| \le \kappa^{-1/2} \epsilon(\epsilon_\varrho, \bar{\epsilon}_f, \bar{\epsilon}_g, \sigma_f, \sigma_g)^{1/2}, x_d \in \Omega_d \Big\}$,
as $t \ge T_1 = \max\{T,\Delta\}$;
and

\item
The mean-square output error satisfies, for $t > 0$,
\[\frac{1}{t} \int^t_0 e^2_1(s) ds \le \frac{1}{t} (c_1 \varpi_0 + c_2) + 2 c_1 \kappa^{-1}
 \epsilon\big(\epsilon_\varrho, \bar{\epsilon}_f, \bar{\epsilon}_g, \sigma_f, \sigma_g\big)
 \]
where
$\varpi_0 =
   \kappa^{-1} \tilde{W}^T_f(0) \Gamma^{-1}_f \tilde{W}_f(0)
 + \kappa^{-1} \tilde{W}^T_g(0) \Gamma^{-1}_g \tilde{W}_g(0)
 + \kappa^{-1} \tau$ $\big({W^\ast_f}^T \Gamma^{-1}_f W^\ast_f + {W^\ast_g}^T \Gamma^{-1}_g W^\ast_g\big)
+ 2 e^2_f(0) \Delta$.
\end{enumerate}
\end{theorem}
{\bf Proof}
Let us consider the integral Lyapunov function candidate,
\[
L(t)= V(t) + 1/2\int^t_{t - \tau} \tilde{W}^T_f(s) \Gamma^{-1}_f \tilde{W}_f(s) ds
+ 1/2 \int^t_{t - \tau} \tilde{W}^T_g(s) \Gamma^{-1}_g \tilde{W}_g(s) ds
\]
with $V(t) = \int^t_0 g(x(s))\frac{d}{ds}$ $\Big(\frac{1}{2}\tilde{e}^2_f(s)\Big)ds$
(the positivity of $V$ can be established,
with the aid of Lemma \ref{Lemma_integral}
and $\tilde{e}_f(0) = 0$).
The derivative of $L$ with respect to time along \re{tilde_e3} is calculated as,
\begin{align}
   \dot{L}
=& - \kappa \tilde{e}^2_f
   - \frac{\tilde{e}^2_f \varrho^2} {\sqrt{\tilde{e}^2_f \varrho^2 + \epsilon^2_\varrho}}\ne
 &
 + \tilde{e}_f
   \Bigg(f(x) - f(x_d)
       + {\tilde{W}_f}^T S_f(x_d)
       + \epsilon_f(x_d)\ne
&\qquad + \nu\left(g(x) - g(x_d) + {\tilde{W}_g}^T S_g(x_d) + \epsilon_g(x_d)\right)\Bigg)\ne
 &
 + \frac{1}{2} \left(\tilde{W}^T_f(t) \Gamma^{-1}_f \tilde{W}_f(t) - \tilde{W}^T_f(t - \tau) \Gamma^{-1}_f \tilde{W}_f(t - \tau)\right)\ne
 &
 + \frac{1}{2} \left(\tilde{W}^T_g(t) \Gamma^{-1}_g \tilde{W}_g(t) - \tilde{W}^T_g(t - \tau) \Gamma^{-1}_g \tilde{W}_g(t - \tau)\right)\nn
\end{align}
According to Assumption \ref{assum2},
\begin{align}
 \dot{L}
\le&
 - \kappa \tilde{e}^2_f
 - \frac{\tilde{e}^2_f \varrho^2} {\sqrt{\tilde{e}^2_f \varrho^2 + \epsilon^2_\varrho}}
 + \tilde{e}_f {\tilde{W}_f}^T S_f
 + \tilde{e}_f \nu {\tilde{W}_g}^T S_g\ne
&
 + \left|\tilde{e}_f\right|
   \left(l_f(x, x_d) + l_g(x, x_d) |\nu| + \bar{\epsilon}_f + \bar{\epsilon}_g |\nu|\right)\ne
 &
 + \frac{1}{2} \left(\tilde{W}^T_f(t) \Gamma^{-1}_f \tilde{W}_f(t) - \tilde{W}^T_f(t - \tau) \Gamma^{-1}_f \tilde{W}_f(t - \tau)\right)\ne
 &
 + \frac{1}{2} \left(\tilde{W}^T_g(t) \Gamma^{-1}_g \tilde{W}_g(t) - \tilde{W}^T_g(t - \tau) \Gamma^{-1}_g \tilde{W}_g(t - \tau)\right)\nn
\end{align}

By virtue of Young's inequality, there is a constant $\eta > 0$ such that
\[- (4 \eta)^{-1} \tilde{e}^2_f + \bar{\epsilon}_f |\tilde{e}_f| \le \eta \bar{\epsilon}^2_f \]
\[- (4 \eta)^{-1} \nu^2 \tilde{e}^2_f + \bar{\epsilon}_g |\tilde{e}_f| |\nu| \le \eta \bar{\epsilon}^2_g\]
It follows from the definition of $\varrho$ in \re{u_DA} that
\begin{align}
\dot{L}
\le&
 - \kappa \tilde{e}^2_f
 + \tilde{e}_f {\tilde{W}_f}^T S_f(x_d)
 + \tilde{e}_f \nu {\tilde{W}_g}^T S_g(x_d)
 - \frac{1 + \nu^2}{4\eta} \tilde{e}^2_f\ne
&
 - \frac{\tilde{e}^2_f \varrho^2} {\sqrt{\tilde{e}^2_f \varrho^2 + \epsilon^2_\varrho}}
 + \left|\tilde{e}_f\right| \varrho
 + \frac{\tilde{e}^2_f}{4 \eta}
 + \frac{\tilde{e}^2_f \nu^2}{4 \eta}
 + \eta \bar{\epsilon}^2_f
 + \eta \bar{\epsilon}^2_g\ne
 &
 + \frac{1}{2} \left(\tilde{W}^T_f(t) \Gamma^{-1}_f \tilde{W}_f(t) - \tilde{W}^T_f(t - \tau) \Gamma^{-1}_f \tilde{W}_f(t - \tau)\right)\ne
 &
 + \frac{1}{2} \left(\tilde{W}^T_g(t) \Gamma^{-1}_g \tilde{W}_g(t) - \tilde{W}^T_g(t - \tau) \Gamma^{-1}_g \tilde{W}_g(t - \tau)\right)\ne
\le&
 - \kappa \tilde{e}^2_f
 + \tilde{e}_f {\tilde{W}_f}^T S_f
 + \tilde{e}_f \nu {\tilde{W}_g}^T S_g
 + \eta \bar{\epsilon}^2_f
 + \eta \bar{\epsilon}^2_g
 + \epsilon_\varrho \ne
 &
 - \tilde{W}^T_f(t) \Gamma^{-1}_f \left(\hat{W}_f(t) - \hat{W}_f(t - \tau)\right)\ne
 &
 - \tilde{W}^T_g(t) \Gamma^{-1}_g \left(\hat{W}_g(t) - \hat{W}_g(t - \tau)\right)\nn
\end{align}
where the inequality,
$- \tilde{e}^2_f \varrho^2 (\tilde{e}^2_f \varrho^2 + \epsilon^2_\varrho)^{-1/2} + \big|\tilde{e}_f\big| \varrho \le \epsilon_\varrho$,
plays an important role,
and
\[\tilde{W}^T_f(t) \Gamma^{-1}_f \tilde{W}_f(t) - \tilde{W}^T_f(t - \tau) \Gamma^{-1}_f \tilde{W}_f(t - \tau) \le - 2 \tilde{W}^T_f(t) \Gamma^{-1}_f \big(\hat{W}_f(t) - \hat{W}_f(t - \tau)\big)
\]
\[\tilde{W}^T_g(t) \Gamma^{-1}_g \tilde{W}_g(t) - \tilde{W}^T_g(t - \tau) \Gamma^{-1}_g \tilde{W}_g(t - \tau) \le - 2 \tilde{W}^T_g(t) \Gamma^{-1}_g \big(\hat{W}_g(t) - \hat{W}_g(t - \tau)\big)
\]
with the help of the algebraic relation $(a-b)^T (a-b) - (a-c)^T (a-c) = (c-b)^T[2(a-b) + (b-c)]$.

To proceed,
applying the weight adaptation laws \re{hatW_fDA} and \re{hatW_gDA} leads to
\begin{align}
\dot{L}
\le&
 - \kappa \tilde{e}^2_f
 + \tilde{e}_f {\tilde{W}_f}^T S_f
 + \tilde{e}_f \nu {\tilde{W}_g}^T S_g
 + \eta \bar{\epsilon}^2_f
 + \eta \bar{\epsilon}^2_g
 + \epsilon_\varrho \ne
 &
 - \tilde{W}^T_f(t) \Gamma^{-1}_f \left(\Gamma_f \tilde{e}_f(t) S_f(x_d) - \sigma_f \Gamma_f \hat{W}_f(t)\right)\ne
 &
 - \tilde{W}^T_g(t) \Gamma^{-1}_g \left(\Gamma_g \tilde{e}_f(t) \nu(t) S_g(x_d)- \sigma_g \Gamma_g \hat{W}_g(t)\right)\ne
=&
 - \kappa \tilde{e}^2_f
 + \sigma_f \tilde{W}^T_f \hat{W}_f
 + \sigma_g \tilde{W}^T_g \hat{W}_g
 + \eta \bar{\epsilon}^2_f
 + \eta \bar{\epsilon}^2_g
 + \epsilon_\varrho\nn
\end{align}

Using the fact that $\sigma_h \tilde{W}^T_h \hat{W}_h = \sigma_h \tilde{W}^T_h W^\ast_h - \sigma_h \tilde{W}^T_h \tilde{W}_h
\le - 2^{-1} \sigma_h \|\tilde{W}_h\|^2 + 2^{-1} \sigma_h \|W^\ast_h\|^2, h \in \{f,g\}$,
it gives rise to
\begin{align}
\dot{L}
\le&
   - \kappa \tilde{e}^2_f
   - \frac{\sigma_f}{2} \|\tilde{W}_f\|^2
   - \frac{\sigma_g}{2} \|\tilde{W}_g\|^2
   + \epsilon\left(\epsilon_\varrho, \bar{\epsilon}_f, \bar{\epsilon}_g, \sigma_f, \sigma_g\right)
\label{dotL}
\end{align}
Define
$\Omega_\epsilon
= \Big\{\tilde{e}_f(t) \big| |\tilde{e}_f(t)| \le \kappa^{-1/2}
 \epsilon \big(\epsilon_\varrho, \bar{\epsilon}_f, \bar{\epsilon}_g, \sigma_f, \sigma_g\big)^{1/2}\Big\}$.
It is clear that the set $\Omega_\epsilon$ is a compact set,
and $\dot{L}$ is negative outside this set.
As such,
$|\tilde{e}_f|$ will decrease and finally converge within the region.
This establishes the UUB property of $\tilde{e}_f$,
implying that there exists the time constant $T$ such that $\tilde{e}_f(t) \in \Omega_\epsilon$ for $t \ge T$.
Similarly,
from \re{dotL},
the UUB property of both $\hat{W}_f$ and $\hat{W}_g$ can also be guaranteed.
Besides,
the boundedness of $e_f(t)$ can be immediately established by
$|e_f(t)| \le |e^\ast_f(t)| + |\tilde{e}_f(t)| \le |e_f(0)| + \kappa^{-1/2}
 \epsilon\big(\epsilon_\varrho, \bar{\epsilon}_f, \bar{\epsilon}_g, \sigma_f, \sigma_g\big)^{1/2}, \forall t \ge T$.
Since $e^\ast_f(t) = 0, \forall t \ge \Delta$,
then $|e_f(t)| \le \kappa^{-1/2} \epsilon\big(\epsilon_\varrho, \bar{\epsilon}_f, \bar{\epsilon}_g, \sigma_f, \sigma_g\big)^{1/2}$, as $t \ge T_1$.
In turn,
$x(t)$ remains in $\Omega_{x}$, for $t \in [T,+\infty)$,
and $x(t) \in \bar{\Omega}_{x}$,  as $t \ge T_1$.
Due to boundedness of $e_f, x_d, \hat{W}_f$ and $\hat{W}_g$,
it is easy from \re{u_DA} to establish the boundedness of the control input, $u$.

Now,
integrating \re{dotL} over $[0,t]$ results in
\begin{align}
   L(t)
\le&
   L(0)
 - \kappa \int^t_0 \tilde{e}^2_f(s) ds
 - \frac{\sigma_f}{2} \int^t_0 \|\tilde{W}_f(s)\|^2 ds\ne
   &
 - \frac{\sigma_g}{2} \int^t_0 \|\tilde{W}_g(s)\|^2 ds
 + \epsilon\left(\epsilon_\varrho, \bar{\epsilon}_f, \bar{\epsilon}_g, \sigma_f, \sigma_g\right) t\nn
\end{align}
Taking into account the non-negativity of $L(t)$ and the fact that $L(0) \in L_\infty$,
one has
\begin{align}
\int^t_0 \tilde{e}^2_f(s) ds
\le&
   \frac{1}{\kappa} \left(L(0) - L(t)\right)
 + \frac{t}{\kappa} \epsilon\left(\epsilon_\varrho, \bar{\epsilon}_f, \bar{\epsilon}_g, \sigma_f, \sigma_g\right)\ne
\le&
   \frac{1}{2 \kappa} \left(\tilde{W}^T_f(0) \Gamma^{-1}_f \tilde{W}_f(0) + \tilde{W}^T_g(0) \Gamma^{-1}_g \tilde{W}_g(0)\right)\ne
 +&\frac{\tau}{2 \kappa} {W^\ast_f}^T \Gamma^{-1}_f W^\ast_f
 + \frac{\tau}{2 \kappa} {W^\ast_g}^T \Gamma^{-1}_g W^\ast_g\ne
 +&\frac{t}{\kappa} \epsilon\left(\epsilon_\varrho, \bar{\epsilon}_f, \bar{\epsilon}_g, \sigma_f, \sigma_g\right)\nn
\end{align}
In addition, utilizing that $|e^\ast_f(t)| \le |e_f(0)|$, for $t \ge 0$,
and $e^\ast_f(t) = 0$, as $t \ge \Delta$,
it can be obtained that
\begin{align}
    \int^t_0 e^2_f(s) ds
\le&2 \int^t_0 \tilde{e}^2_f(s) ds + 2 \int^\Delta_0 {e^\ast_f}^2(s) ds\ne
\le&
   \frac{1}{\kappa} \left(\tilde{W}^T_f(0) \Gamma^{-1}_f \tilde{W}_f(0) + \tilde{W}^T_g(0) \Gamma^{-1}_g \tilde{W}_g(0)\right)\ne
 +&\frac{\tau}{\kappa} {W^\ast_f}^T \Gamma^{-1}_f W^\ast_f
 + \frac{\tau}{\kappa} {W^\ast_g}^T \Gamma^{-1}_g W^\ast_g\ne
 +&\frac{2 t}{\kappa} \epsilon\left(\epsilon_\varrho, \bar{\epsilon}_f, \bar{\epsilon}_g, \sigma_f, \sigma_g\right)
 + 2 e^2_f(0) \Delta\nn
\end{align}
According to the relationship between $e_1$ and $e_f$,
the mean-square output error $\frac{1}{t} \int^t_0 e^2_1(s) ds$ can be established.
This completes the proof.
\QED

\begin{remark}
The mean-square error given by Theorem \ref{thm.2} reveals the transient response of the closed-loop system,
which reflects the average tracking performance over the whole control process.
Furthermore,
taking the limit leads to $\limsup_{t \rightarrow \infty} \frac{1}{t} \int^t_0 e^2_1(s) ds \le 2 c_1 \kappa^{-1}
 \epsilon\big(\epsilon_\varrho, \bar{\epsilon}_f, \bar{\epsilon}_g, \sigma_f, \sigma_g\big)$,
implying that
the output error $e_1(t)$ will converge to the neighborhood of the origin in the sense of mean-square,
and the radius of the neighborhood is adjustable.
\end{remark}

\begin{corollary}
\label{cor_filtered_error1}
Let the controller \re{u_DA} together with the adaptation laws \re{hatW_fDA} and \re{hatW_gDA}
be applied to system \re{sys},
under Assumptions \ref{assum1} and \ref{assum2}.
Then,
the filtered error satisfies that
\[
\limsup_{t \rightarrow \infty} \int^t_{t - \tau} e^2_f (s) ds \le \tau \kappa^{-1}
\epsilon \big(\epsilon_\varrho, \bar{\epsilon}_f, \bar{\epsilon}_g, \sigma_f, \sigma_g\big)
\]
Moreover,
\[\limsup_{t \rightarrow \infty} \int^t_{t - \tau} \|\tilde{W}_f(s)\|^2 ds
\le
2 \sigma^{-1}_f \tau\epsilon\big(\epsilon_\varrho, \bar{\epsilon}_f, \bar{\epsilon}_g, \sigma_f, \sigma_g\big)\]
\[
\limsup_{t \rightarrow \infty} \int^t_{t - \tau} \|\tilde{W}_g(s)\|^2 ds
\le
2 \sigma^{-1}_g \tau \epsilon\big(\epsilon_\varrho, \bar{\epsilon}_f, \bar{\epsilon}_g, \sigma_f, \sigma_g\big)
\]
\end{corollary}
{Proof}
As in the proof of Theorem \ref{thm.2},
again integrating \re{dotL} over $[t - \tau, t]$ gives rise to
\begin{align}
  L(t) - L(t - \tau)
\le&
 - \kappa \int^t_{t - \tau} \tilde{e}^2_f(s) ds
 - \frac{\sigma_f}{2} \int^t_{t - \tau} \|\tilde{W}_f(s)\|^2 ds\ne
   &
 - \frac{\sigma_g}{2} \int^t_{t - \tau} \|\tilde{W}_g(s)\|^2 ds
 + \epsilon\left(\epsilon_\varrho, \bar{\epsilon}_f, \bar{\epsilon}_g, \sigma_f, \sigma_g\right) \tau \nn
\end{align}
For the case where $L(t_0), t_0 \in [0, \tau]$ is finite,
in the light of Lemma \ref{keylem},
the integral term $\int^t_{t - \tau} \tilde{e}^2_f (s) ds$ could enter the specified bound $\tau \kappa^{-1} \epsilon \big(\epsilon_\varrho, \bar{\epsilon}_f, \bar{\epsilon}_g, \sigma_f, \sigma_g\big)$,
as $t \rightarrow +\infty$.
Since $\lim_{t \rightarrow \infty} e^\ast_f(t) = 0$,
the filtered error satisfies
\begin{align}
\limsup_{t \rightarrow \infty} \int^t_{t - \tau} e^2_f (s) ds
\le&
  \frac{\tau}{\kappa} \epsilon\left(\epsilon_\varrho, \bar{\epsilon}_f, \bar{\epsilon}_g, \sigma_f, \sigma_g\right)
  \label{limsup_ef1}
\end{align}
In order to obtain the above convergence performance of the filtered error,
we have to consider the finiteness of $L(t_0)$.
It follows from \re{dotL} that $\dot{L}(t_0) \le \epsilon \big(\epsilon_\varrho, \bar{\epsilon}_f, \bar{\epsilon}_g, \sigma_f, \sigma_g\big)$.
Since $W^\ast_f$ and $W^\ast_g$ are bounded,
$L(t_0)$ is bounded on the interval $[0,\tau]$.
Similarly,
the following relationships can be deduced,
by invoking Lemma \ref{keylem},
\begin{align}
   \limsup_{t \rightarrow \infty} \int^t_{t - \tau} \|\tilde{W}_f(s)\|^2 ds
\le&
   \frac{2 \tau}{\sigma_f} \epsilon\left(\epsilon_\varrho, \bar{\epsilon}_f, \bar{\epsilon}_g, \sigma_f, \sigma_g\right)
\label{limsup_Wf1}\\
   \limsup_{t \rightarrow \infty} \int^t_{t - \tau} \|\tilde{W}_g(s)\|^2 ds
\le&
   \frac{2 \tau}{\sigma_g} \epsilon\left(\epsilon_\varrho, \bar{\epsilon}_f, \bar{\epsilon}_g, \sigma_f, \sigma_g\right)
\label{limsup_Wg1}
\end{align}
This completes the proof.
\QED

Corollary \ref{cor_filtered_error1} establishes the asymptotic convergence of the filtered error
in the sense of $L_2$ over $[t - \tau, t]$, as $t \rightarrow \infty$,
and in turn assures the asymptotic convergence of the tracking error, where Lemma \ref{keylem} plays a key role to finalize the analysis.

With the aid of the proposed DA-based adaptive neural controller,
theoretical results on the boundedness of variables and the convergence of the tracking error are established
in Theorem \ref{thm.2} and Corollary \ref{cor_filtered_error1}.
Assumption \ref{assum2} is one of the essential aspects to achieve these and,
in order to fulfill it,
there is usually a necessity to know the nonlinear functions $l_f(x, x_d)$ and $l_g(x, x_d)$.
As a matter of fact,
this requirement is not difficult to satisfy,
as long as the pre-designed functions $l_f(x, x_d)$ and $l_g(x, x_d)$ are sufficiently large.
In order to relax this condition to facilitate the realisation of the control design,
the following assumption is made,
taking into account and making adjustments for Assumption \ref{assum2},
\begin{assumption}
\label{assum_Lipschitz}
The uncertain nonlinearities, $f(x)$ and $g(x)$, involved in the system dynamic undertaken,
satisfy the following conditions
\begin{align}
    |f(\chi_1) - f(\chi_2)| \le& l_f \|\chi_1 - \chi_2\| \hbar_f(\chi_1, \chi_2)
\label{l_f} \\
    |g(\chi_1) - g(\chi_2)| \le& l_g \|\chi_1 - \chi_2\| \hbar_g(\chi_1, \chi_2)
\label{l_g1}
\end{align}
$\forall \chi_1, \chi_2 \in R^n$,
where $l_f$ and $l_g$ are positive constants that are not required to be known,
and $\hbar_h(\cdot,\cdot)$ is a known and nonnegative continuous function, for $h \in \{f,g\}$.
\end{assumption}

Define $\hat{l}_f$ and $\hat{l}_g$ as the estimates of the unknown parameters $l_f$ and $l_g$,
respectively,
and let $\tilde{l}_f = l_f - \hat{l}_f$ and $\tilde{l}_g = l_g - \hat{l}_g$ be the parameter estimation errors.
In order to update $\hat{l}_f$ and $\hat{l}_g$,
the following learning laws, designed with refer to the adaptive laws \re{hatW_fDA} and \re{hatW_gDA},
are adopted,
\begin{align}
   (1 + \check{\sigma}_f \gamma_f) \hat{l}_f(t)
=& \hat{l}_f(t - \tau)
+ \gamma_f |\tilde{e}_f(t)| \|e(t)\| \hbar_f(x, x_d)
\label{hatl_f} \\
   (1 + \check{\sigma}_g \gamma_g) \hat{l}_g(t)
=& \hat{l}_g(t - \tau)
+ \gamma_g |\nu(t)| |\tilde{e}_f(t)| \|e(t)\| \hbar_g(x, x_d)
\label{hatl_g}
\end{align}
for which a adaptive neural controller of the following form:
\begin{align}
u =& - \tilde{e}_f \left(\frac{\varsigma^2 \|e\|^2}{\sqrt{\varsigma^2 \tilde{e}^2_f \|e\|^2 + \epsilon^2_\varsigma}} + \kappa + \frac{1}{4 \eta_f} + \frac{\nu^2}{4 \eta_g}\right)
    - \hat{W}^T_f S_f(x_d)
    - \hat{W}^T_g S_g(x_d) \nu
\label{u_DA2}
\end{align}
is applicable,
where $\varsigma = \hat{l}_f \hbar_f(x, x_d) + \hat{l}_g |\nu| \hbar_g(x, x_d)$,
and $\check{\sigma}_f$, $\check{\sigma}_g$, $\epsilon_\varsigma$, $\eta_f$, $\eta_g$, $\gamma_f$, and $\gamma_g$
are positive adjustable parameters.

\begin{theorem}
\label{thm.3}
For system \re{sys} with Assumptions \ref{assum1} and \ref{assum_Lipschitz} being satisfied,
the adaptive neural controller \re{u_DA2} is adopted,
together with adaptation laws \re{hatW_fDA}, \re{hatW_gDA}, \re{hatl_f} and \re{hatl_g}.
Then
\begin{enumerate}
\item
all the variables in the closed-loop system are bounded;

\item
the state $x$ remains in $\check{\Omega}_x = \Big\{x(t) \big| |e_f(t)| \le |e_f(0)|$ $+ \kappa^{-1/2} \epsilon \big(\epsilon_\varsigma, \bar{\epsilon}_f, \bar{\epsilon}_g, \sigma_f, \sigma_g, \check{\sigma}_f, \check{\sigma}_g\big)^{1/2}, x_d \in \Omega_d \Big\}$
for $t \ge T$;
and $x(t) \in \check{\bar{\Omega}}_x = \Big\{x(t) \big| |e_f(t)| \le \kappa^{-1/2}\epsilon \big(\epsilon_\varsigma,\bar{\epsilon}_f, \bar{\epsilon}_g, $ $\sigma_f, \sigma_g, \check{\sigma}_f, \check{\sigma}_g\big)^{1/2}, x_d \in \Omega_d \Big\}$,
as $t \ge T_1 = \max\{T,\Delta\}$,
where $\epsilon \big(\epsilon_\varsigma, \bar{\epsilon}_f, \bar{\epsilon}_g, \sigma_f, \sigma_g, \check{\sigma}_f, \check{\sigma}_g\big)
= \epsilon_\varsigma
+ \eta_f \bar{\epsilon}^2_f
+ \eta_g \bar{\epsilon}^2_g
+ 2^{-1} \sigma_f \|W^\ast_f\|^2
+ 2^{-1} \sigma_g \|W^\ast_g\|^2
+ 2^{-1} \check{\sigma}_f l^2_f
+ 2^{-1} \check{\sigma}_g l^2_g$;
and

\item
the mean-square output error satisfies that
\[
\frac{1}{t} \int^t_0 e^2_1(s) ds \le \frac{1}{t} (c_1 \check{\varpi}_0 + c_2) + 2 c_1 \kappa^{-1} \epsilon \big(\epsilon_\varsigma,\bar{\epsilon}_f, \bar{\epsilon}_g, \sigma_f, \sigma_g, \check{\sigma}_f, \check{\sigma}_g\big)
\]
where $\check{\varpi}_0 =
   \kappa^{-1} (\tilde{W}^T_f(0) \Gamma^{-1}_f \tilde{W}_f(0) + \tilde{W}^T_g(0) \Gamma^{-1}_g \tilde{W}_g(0))
 + \kappa^{-1} \tau ({W^\ast_f}^T \Gamma^{-1}_f W^\ast_f + {W^\ast_g}^T \Gamma^{-1}_g W^\ast_g)
 + (\kappa \gamma_f)^{-1} \tilde{l}^2_f(0)
 + (\kappa \gamma_g)^{-1} \tilde{l}^2_g(0)
 + \tau (\kappa \gamma_f)^{-1} l^2_f
 + \tau (\kappa \gamma_g)^{-1} l^2_g
 + 2 e^2_f(0) \Delta$.
\end{enumerate}
\end{theorem}
{\bf Proof}
Let us choose the integral Lyapunov function candidate,
$V(t) = \int^t_0 g(x(s)) \frac{d}{ds} \Big(\frac{1}{2}\tilde{e}^2_f(s)\Big) ds$.
Under \re{l_f} and \re{l_g1},
the derivative of $V$ with respect to time along \re{tilde_e3} can be calculated as
\begin{align}
   \dot{V}
=& \tilde{e}_f \Big(f(x) - f(x_d) + {W^\ast_f}^T S_f(x_d) + \epsilon_f(x_d) + u\ne
   &+\nu\left(g(x) - g(x_d) + {W^\ast_g}^T S_g(x_d) + \epsilon_g(x_d)\right)\Big)\ne
\le&
   |\tilde{e}_f| \left(l_f \|e\| \hbar_f + l_g |\nu| \|e\| \hbar_g + \bar{\epsilon}_f + |\nu| \bar{\epsilon}_g\right)\ne
   &
 + \tilde{e}_f \left({W^\ast_f}^T S_f(x_d) + \nu {W^\ast_g}^T S_g(x_d) + u\right)\nn
\end{align}

By means of Young's inequality,
one obtains that $\bar{\epsilon}_f |\tilde{e}_f| \le (4 \eta_f)^{-1} \tilde{e}^2_f + \eta_f \bar{\epsilon}^2_f$
and $\bar{\epsilon}_g |\tilde{e}_f| |\nu| \le (4 \eta_g)^{-1} \nu^2 \tilde{e}^2_f + \eta_g \bar{\epsilon}^2_g$, which implies
\begin{align}
   \dot{V}
\le&
   \tilde{e}_f \left({W^\ast_f}^T S_f(x_d) + \nu {W^\ast_g}^T S_g(x_d) + u\right)\ne
   &
 + \tilde{l}_f |\tilde{e}_f| \|e\| \hbar_f
 + \tilde{l}_g |\tilde{e}_f| |\nu| \|e\| \hbar_g
 + \varsigma |\tilde{e}_f| \|e\|\ne
   &
 + \left(\frac{1}{4 \eta_f} + \frac{\nu^2}{4 \eta_g}\right) \tilde{e}^2_f
 + \eta_f \bar{\epsilon}^2_f
 + \eta_g \bar{\epsilon}^2_g
\label{V_dot_1}
\end{align}
Substituting \re{u_DA2} into \re{V_dot_1} results in
\begin{align}
   \dot{V}
\le&
 - \kappa \tilde{e}^2_f
 - \frac{1}{4 \eta_f} \tilde{e}^2_f
 - \frac{\nu^2}{4 \eta_g} \tilde{e}^2_f
 + \left(\frac{1}{4 \eta_f} + \frac{\nu^2}{4 \eta_g}\right) \tilde{e}^2_f \ne
 &
 + \tilde{e}_f \left({W^\ast_f}^T S_f + \nu {W^\ast_g}^T S_g - \hat{W}^T_f S_f - \nu \hat{W}^T_g S_g\right)\ne
   &
 + \tilde{l}_f |\tilde{e}_f| \|e\| \hbar_f
 + \tilde{l}_g |\tilde{e}_f| |\nu| \|e\| \hbar_g
 + \eta_f \bar{\epsilon}^2_f
 + \eta_g \bar{\epsilon}^2_g\ne
   &
 - \frac{\varsigma^2 \tilde{e}^2_f \|e\|^2}{\sqrt{\varsigma^2 \tilde{e}^2_f \|e\|^2 + \epsilon^2_\varsigma}}
 + \varsigma |\tilde{e}_f| \|e\|\ne
\le&
 - \kappa \tilde{e}^2_f
 + \tilde{e}_f \tilde{W}^T_f S_f(x_d)
 + \tilde{e}_f \nu \tilde{W}^T_g S_g(x_d)
 + \epsilon_\varsigma\ne
   &
 + \tilde{l}_f |\tilde{e}_f| \|e\| \hbar_f(x, x_d)
 + \tilde{l}_g |\tilde{e}_f| |\nu| \|e\| \hbar_g(x, x_d)\ne
 &
 + \eta_f \bar{\epsilon}^2_f
 + \eta_g \bar{\epsilon}^2_g
 \label{V_dot_2}
\end{align}
where $- \varsigma^2 \tilde{e}^2_f \|e\|^2 (\varsigma^2 \tilde{e}^2_f \|e\|^2 + \epsilon^2_\varsigma)^{-1/2} \le
 - |\varsigma| |\tilde{e}_f| \|e\| + \epsilon_\varsigma$.

To proceed,
consider the Lyapunov-Krasovskii function candidate,
\bea
\check{L} (t) &=& V(t)
   + 1/2 \int^t_{t - \tau} \tilde{W}^T_f(s) \Gamma^{-1}_f \tilde{W}_f(s) ds
   + 1/2 \int^t_{t - \tau} \tilde{W}^T_g(s) \Gamma^{-1}_g \tilde{W}_g(s) ds \ne &&
   + 1/2 \gamma_f^{-1} \int^t_{t - \tau} \tilde{l}^2_f(s) ds
   + 1/2 \gamma_g^{-1}\int^t_{t - \tau} \tilde{l}^2_g(s) ds
\eea
Making use of \re{V_dot_2},
its time derivative can be expressed as
\begin{align}
   \dot{\check{L}}
\le&
 - \kappa \tilde{e}^2_f
 + \tilde{e}_f \tilde{W}^T_f S_f(x_d)
 + \tilde{e}_f \nu \tilde{W}^T_g S_g(x_d)
 + \epsilon_\varsigma\ne
   &
 + \tilde{l}_f |\tilde{e}_f| \|e\| \hbar_f
 + \tilde{l}_g |\tilde{e}_f| |\nu| \|e\| \hbar_g
 + \eta_f \bar{\epsilon}^2_f
 + \eta_g \bar{\epsilon}^2_g\ne
 &
 + \frac{1}{2} \left(\tilde{W}^T_f(t) \Gamma^{-1}_f \tilde{W}_f(t) - \tilde{W}^T_f(t - \tau) \Gamma^{-1}_f \tilde{W}_f(t - \tau)\right)\ne
 &
 + \frac{1}{2} \left(\tilde{W}^T_g(t) \Gamma^{-1}_g \tilde{W}_g(t) - \tilde{W}^T_g(t - \tau) \Gamma^{-1}_g \tilde{W}_g(t - \tau)\right)\ne
 &
 + \frac{1}{2 \gamma_f} \left(\tilde{l}^2_f(t) - \tilde{l}^2_f(t - \tau)\right)
 + \frac{1}{2 \gamma_g} \left(\tilde{l}^2_g(t) - \tilde{l}^2_g(t - \tau)\right)\nn
\end{align}

Using the facts that
$\tilde{W}^T_f(t) \Gamma^{-1}_f \tilde{W}_f(t) - \tilde{W}^T_f(t - \tau) \Gamma^{-1}_f$ $\tilde{W}_f(t - \tau)
\le - 2 \tilde{W}^T_f(t) \Gamma^{-1}_f \big(\hat{W}_f(t) - \hat{W}_f(t - \tau)\big),$ $
\tilde{W}^T_g(t) \Gamma^{-1}_g$ $\tilde{W}_g(t) - \tilde{W}^T_g(t - \tau) \Gamma^{-1}_g \tilde{W}_g(t - \tau)
\le - 2 \tilde{W}^T_g(t)\Gamma^{-1}_g $ $\big(\hat{W}_g(t) - \hat{W}_g(t - \tau)\big),
\tilde{l}^2_f(t) - \tilde{l}^2_f(t - \tau) \le -2 \tilde{l}_f(t) \big(\hat{l}_f(t) - \hat{l}_f(t - \tau)\big)$
and
$\tilde{l}^2_g(t) - \tilde{l}^2_g(t - \tau) \le -2 \tilde{l}_g(t) \big(\hat{l}_g(t) - \hat{l}_g(t - \tau)\big)$,
it follows that
\begin{align}
\dot{\check{L}}
\le&
 - \kappa \tilde{e}^2_f
 + \tilde{e}_f \tilde{W}^T_f S_f(x_d)
 + \tilde{e}_f \nu \tilde{W}^T_g S_g(x_d)
 + \epsilon_\varsigma\ne
   &
 + \tilde{l}_f |\tilde{e}_f| \|e\| \hbar_f
 + \tilde{l}_g |\tilde{e}_f| |\nu| \|e\| \hbar_g
 + \eta_f \bar{\epsilon}^2_f
 + \eta_g \bar{\epsilon}^2_g\ne
 &
 - \tilde{W}^T_f(t) \Gamma^{-1}_f \left(\hat{W}_f(t) - \hat{W}_f(t - \tau)\right)\ne
 &
 - \tilde{W}^T_g(t) \Gamma^{-1}_g \left(\hat{W}_g(t) - \hat{W}_g(t - \tau)\right)\ne
 &
 - \frac{1}{\gamma_f} \tilde{l}_f(t) \left(\hat{l}_f(t) - \hat{l}_f(t - \tau)\right)\ne
 &
 - \frac{1}{\gamma_g} \tilde{l}_g(t) \left(\hat{l}_g(t) - \hat{l}_g(t - \tau)\right)\nn
\end{align}
Further,
using \re{hatW_fDA}, \re{hatW_gDA}, \re{hatl_f} and \re{hatl_g} implies that
\begin{align}
\dot{\check{L}}
\le&
 - \kappa \tilde{e}^2_f
 + \epsilon_\varsigma
 + \sigma_f \tilde{W}^T_f \hat{W}_f
 + \sigma_g \tilde{W}^T_g \hat{W}_g\ne
   &
 + \check{\sigma}_f \tilde{l}_f \hat{l}_f
 + \check{\sigma}_g \tilde{l}_g \hat{l}_g
 + \eta_f \bar{\epsilon}^2_f
 + \eta_g \bar{\epsilon}^2_g\ne
\le&
 - \kappa \tilde{e}^2_f
 - \frac{\sigma_f}{2} \|\tilde{W}_f\|^2
 - \frac{\sigma_g}{2} \|\tilde{W}_g\|^2
 - \frac{\check{\sigma}_f}{2} \tilde{l}^2_f
 - \frac{\check{\sigma}_g}{2} \tilde{l}^2_g\ne
 &
 + \epsilon \left(\epsilon_\varsigma, \bar{\epsilon}_f, \bar{\epsilon}_g, \sigma_f, \sigma_g, \check{\sigma}_f, \check{\sigma}_g\right)
\label{L_dot}
\end{align}
where
$\sigma_f \tilde{W}^T_f \hat{W}_f \le - 2^{-1} \sigma_f \big(\|\tilde{W}_f\|^2 - \|W^\ast_f\|^2\big)$,
$\sigma_g \tilde{W}^T_g \hat{W}_g \le - 2^{-1} \sigma_g  \big(\|\tilde{W}_g\|^2 - \|W^\ast_g\|^2\big)$,
$\check{\sigma}_f \tilde{l}_f \hat{l}_f \le - 2^{-1} \check{\sigma}_f \big(\tilde{l}^2_f - l^2_f\big)$
and $\check{\sigma}_g \tilde{l}_g \hat{l}_g \le - 2^{-1} \check{\sigma}_g \big(\tilde{l}^2_g - l^2_g\big)$
are used.

Similar to the proof of Theorem \ref{thm.2},
from \re{L_dot},
it is easy to establish the UUB property of $\tilde{e}_f$, $\hat{W}_f$, $\hat{W}_g$, $\hat{l}_f$ and $\hat{l}_g$.
In addition,
there exists a time constant $T$ such that $x(t)$ remains within the compact set $\check{\Omega}_x$ as $t \ge T$.
Similarly,
$x(t)$ remains within the compact set $\check{\bar{\Omega}}_x$ as $t \ge T_1$,
due to the fact that $e^\ast_f(t) = 0, \forall t \ge \Delta$.
In accordance with the boundedness of $\tilde{e}_f$,
one can obtain that $e_f \in L_\infty$ and $e \in L_\infty$
and in turn the system state $x$ is finite.
It is easy to prove that the control input $u$ (see \re{u_DA2}) is bounded,
by exploiting the boundedness of other variables.
Then,
the boundedness of all the internal signals is established.

Similarly,
integrating \re{L_dot} from $0$ to $t$ yields
\begin{align}
\int^t_0 \tilde{e}^2_f(s) ds
\le&
   \frac{1}{2 \kappa} \tilde{W}^T_f(0) \Gamma^{-1}_f \tilde{W}_f(0)
 + \frac{\tau}{2 \kappa} {W^\ast_f}^T \Gamma^{-1}_f W^\ast_f\ne
 &
 + \frac{1}{2 \kappa} \tilde{W}^T_g(0) \Gamma^{-1}_g \tilde{W}_g(0)
 + \frac{\tau}{2 \kappa} {W^\ast_g}^T \Gamma^{-1}_g W^\ast_g\ne
 &
 + \frac{1}{2 \kappa \gamma_f} \tilde{l}^2_f(0)
 + \frac{1}{2 \kappa \gamma_g} \tilde{l}^2_g(0)
 + \frac{\tau}{2 \kappa \gamma_f} l^2_f\ne
 &
 + \frac{\tau}{2 \kappa \gamma_g} l^2_g
 + \frac{t}{\kappa}
   \epsilon \left(\epsilon_\varsigma, \bar{\epsilon}_f, \bar{\epsilon}_g, \sigma_f, \sigma_g, \check{\sigma}_f, \check{\sigma}_g\right)\nn
\end{align}
It can be further concluded that $\int^t_0 e^2_f(s) ds
\le 2 \int^t_0 \tilde{e}^2_f(s) ds
  + 2 \int^t_0 {e^\ast_f}^2(s) ds
\le \kappa^{-1} \big(\tilde{W}^T_f(0) \Gamma^{-1}_f \tilde{W}_f(0) + \tilde{W}^T_g(0) \Gamma^{-1}_g\tilde{W}_g(0)\big)$ $
  + \kappa^{-1} \tau \big({W^\ast_f}^T \Gamma^{-1}_f W^\ast_f + {W^\ast_g}^T \Gamma^{-1}_g W^\ast_g\big)
  + (\kappa \gamma_f)^{-1} \tilde{l}^2_f(0)
  + (\kappa \gamma_g)^{-1} \tilde{l}^2_g(0)
  + (\kappa \gamma_f)^{-1} \tau l^2_f
  + (\kappa \gamma_g)^{-1} \tau l^2_g
  + 2 \kappa^{-1} t
   \epsilon \big(\epsilon_\varsigma, \bar{\epsilon}_f,\bar{\epsilon}_g,$ $ \sigma_f,\sigma_g, \check{\sigma}_f,\check{\sigma}_g\big)
  + 2 e^2_f(0) \Delta$,
where $\int^t_0 {e^\ast_f}^2(s) ds \le \int^\Delta_0 {e^\ast_f}^2(s)$ $ ds \le \int^\Delta_0e^2_f(0) ds$.
By virtue of the relationship between $e_1$ and $e_f$,
the conclusions about the mean-square output error, $\frac{1}{t} \int^t_0 e^2_1(s) ds$,
follow with similar lines to those in the proof for Theorem \ref{thm.2}.
\QED

Furthermore,
again integrating \re{L_dot} over $[t - \tau,t]$ leads to
\begin{align}
   \check{L}(t) - \check{L}(t - \tau)
\le&
 - \kappa \int^t_{t - \tau} \tilde{e}^2_f (s) ds
 - \frac{\sigma_f}{2} \int^t_{t - \tau} \|\tilde{W}_f(s)\|^2 ds
 - \frac{\sigma_g}{2} \int^t_{t - \tau} \|\tilde{W}_g(s)\|^2 ds\ne
 &
 - \frac{\check{\sigma}_f}{2} \int^t_{t - \tau} \tilde{l}^2_f(s) ds
 - \frac{\check{\sigma}_g}{2} \int^t_{t - \tau} \tilde{l}^2_g(s) ds
 + \epsilon \left(\epsilon_\varsigma, \bar{\epsilon}_f, \bar{\epsilon}_g, \sigma_f, \sigma_g, \check{\sigma}_f, \check{\sigma}_g\right) \tau\nn
\end{align}
Since $\lim_{t \rightarrow \infty} e^\ast_f(t) = 0$,
and $\check{L}(t_0) \in L_\infty, \forall t_0 \in [0, \tau]$, due to the fact that
 \[
 \dot{\check{L}}(t_0)
\le \epsilon \big(\epsilon_\varsigma, \bar{\epsilon}_f, \bar{\epsilon}_g, \sigma_f, \sigma_g, \check{\sigma}_f, \check{\sigma}_g\big)
< + \infty)
\]

The convergence conclusions are summarized in the following corollary,
by appealing to Lemma \ref{keylem}.

\begin{corollary}
\label{cor_filtered_error2}
For  system \re{sys}
under Assumptions \ref{assum1} and \ref{assum_Lipschitz},
the adaptive neural control law \re{u_DA2} and the adaptation laws \re{hatW_fDA}, \re{hatW_gDA}, \re{hatl_f} and \re{hatl_g}
can guarantee that

\bea
&& \limsup_{t \rightarrow \infty} \int^t_{t - \tau} e^2_f (s) ds \le \kappa^{-1} \tau
   \epsilon \big(\epsilon_\varsigma, \bar{\epsilon}_f, \bar{\epsilon}_g, \sigma_f, \sigma_g, \check{\sigma}_f, \check{\sigma}_g\big) \ne
&&\limsup_{t \rightarrow \infty} \int^t_{t - \tau} \|\tilde{W}_f(s)\|^2 ds
\le
2 \sigma^{-1}_f \tau
\epsilon \big(\epsilon_\varsigma, \bar{\epsilon}_f, \bar{\epsilon}_g, \sigma_f, \sigma_g, \check{\sigma}_f, \check{\sigma}_g\big)\ne
&& \limsup_{t \rightarrow \infty} \int^t_{t - \tau} \|\tilde{W}_g(s)\|^2 ds
\le
2 \sigma^{-1}_g \tau
\epsilon \big(\epsilon_\varsigma, \bar{\epsilon}_f, \bar{\epsilon}_g, \sigma_f, \sigma_g, \check{\sigma}_f, \check{\sigma}_g\big) \ne
&& \limsup_{t \rightarrow \infty} \int^t_{t - \tau} \tilde{l}^2_f(s) ds
\le
2 \check{\sigma}^{-1}_f \tau
\epsilon \big(\epsilon_\varsigma, \bar{\epsilon}_f, \bar{\epsilon}_g, \sigma_f, \sigma_g, \check{\sigma}_f, \check{\sigma}_g\big)
\ne
&&\limsup_{t \rightarrow \infty} \int^t_{t - \tau} \tilde{l}^2_g(s) ds
\le
2 \check{\sigma}^{-1}_g \tau
\epsilon \big(\epsilon_\varsigma, \bar{\epsilon}_f, \bar{\epsilon}_g, \sigma_f, \sigma_g, \check{\sigma}_f, \check{\sigma}_g\big) \nn
\eea
\end{corollary}

\begin{remark}
\label{remark}
From Theorems \ref{thm.2} and \ref{thm.3} and Corollaries \ref{cor_filtered_error1} and \ref{cor_filtered_error2},
the obtained results about the steady-state and transient responses hold as long as $x_d\in \Omega_d$,
whereas $x \in \Omega_x$ is not required.
The incremental adaptive control of error system \re{tilde_e3}, 
the input elements of the adopted neural networks only relying on the desired trajectory given in advance, is shown to work well.
As such,
the compact set $\Omega_d$ exists,
and in turn the DA approach offers an efficient way to meet the compact set prerequisite for the NN training.
\end{remark}

\section{Conclusion}
\label{sect.conclusion}
This paper presents
the desired approximation approach
to the design of adaptive neural controllers,
where the developed approximation strategy
requires only the desired state signal
for the input of the neural network, but does
not use the practical one.
As a result,
for the controller design, the assumption on
the existence of the compact set of practical state
can be avoided,
and both the approximation and control performance can be improved.
The proposed integral Lyapunov function,
together with the error-tracking technique,
is shown to be efficient
in the treatment of the state-dependent control gain,
such that the controller design can be simplified significantly.
It has been shown that the control scheme works well to assure that,
for any bounded initial conditions including system states and weight estimates,
the tracking error will converge to an adjustable set,
while the boundedness of all variables in the closed-loop system is guaranteed.
Theoretical results of establishing transient performance of the output error have been presented,
which demonstrate effectiveness of the presented adaptive neural control schemes.







{\small
\begin{thebibliography}{99}

%
%
%
%
%
%
\bibitem{krstic95}
M. Krstic, I. Kanellakopoulos, and P. V. Kokotovic,
\emph{Nonlinear and Adaptive Control Design}.
New York, NY: Wiley, 1995.

%


\bibitem{ioannon06}
P.~A.~Ioannon and B.~Fidan,
\emph{Adaptive Control Tutorial}.
Philadelphia, PA: SIAM, 2006.




\bibitem{Ge01}
S. S. Ge, C. C. Hang, T. H. Lee, and T. Zhang,
\emph{Stable Adaptive Neural Network Control}.
Boston, MA, USA: Kluwer, 2001.

\bibitem{Spooner02}
J. Spooner, M. Maggiore, R. Ordonez, and K. Passino,
\emph{Stable Adaptive Control and Estimation for Nonlinear Systems: Neural and Fuzzy Approximator Techniques}.
New York, USA: Wiley, 2002.

\bibitem{Farrell06}
J. A. Farrell and M. M. Polycarpou,
\emph{Adaptive Approximation Based Control: Unifying Neural, Fuzzy and Tranditional Adaptive Approximation Approaches}.
Hoboken, NJ, USA: Wiley, 2006.



\bibitem{Narendra90}
K. S. Narendra and K. Parthasarathy,
``Identification and control of dynamical systems using neural networks,''
\emph{IEEE Trans. Neural Netw.},
vol. 1, no. 1, pp. 4-27, 1990.

\bibitem{Sanner92}
R. M. Sanner and J. J. E. Slotine,
``Gaussian networks for direct adaptive control,''
\emph{IEEE Trans. Neural Netw.},
vol. 3, no. 6, pp. 837-863, 1992.

\bibitem{Sadegh93}
N. Sadegh,
``A perceptron network for functional identification and control of nonlinear systems,''
\emph{IEEE Trans. Neural Netw.},
vol. 4, no. 6, pp. 982-988, Nov. 1993.

\bibitem{Zhang00a}
T. Zhang, S. S. Ge, and C. C. Hang,
``Stable adaptive control for a class of nonlinear systems using a modified Lyapunov function,''
\emph{IEEE Trans. Autom. Control},
vol. 45, no. 1, pp. 129-132, Jan. 2000.



\bibitem{sun06}
M. Sun and S. S. Ge,
``Adaptive repetitive control for a class of nonlinearly parametrized systems,''
\emph{IEEE Trans. Autom. Control},
vol. 51, no. 10, pp. 1684-1688, Oct. 2006.

%
%
%














\bibitem{Volyanskyy09}
K. Y. Volyanskyy, W. M. Haddad and A. J. Calise,
``A new neuroadaptive control architecture for nonlinear uncertain dynamical systems: beyond $\sigma$- and $e$-modifications,''
\emph{IEEE Trans. Neural Netw.},
vol. 20, no. 11, pp. 1707-1723, Nov. 2009.

\bibitem{WangW17}
W. Wang, C. Wen and J. Huang,
``Distributed adaptive asymptotically consensus tracking control of nonlinear multi-agent systems with unknown parameters and uncertain disturbances,''
\emph{Automatica},
vol. 77, pp. 133-142, 2017.


\bibitem{Ge02}
S. S. Ge and C. Wang,
``Adaptive NN control of uncertain nonlinear pure-feedback systems,''
\emph{Automatica},
vol. 38, no. 4, pp. 671-682, 2002.

\bibitem{Wang02}
D. Wang and J. Huang,
``Adaptive neural network control for a class of uncertain nonlinear systems in pure-feedback form,''
\emph{Automatica},
vol. 38, no. 8, pp. 1365-1372, 2002.

\bibitem{ChenB16}
B. Chen, H. Zhang, and C. Lin,
``Observer-based adaptive neural network control for nonlinear systems in nonstrict-feedback form,''
\emph{IEEE Trans. Neural Netw. Learn. Syst.},
vol. 27, no. 1, pp. 89-98, Jan. 2016.



\bibitem{Tee09}
K. P. Tee, S. S. Ge, and E. H. Tay,
``Barrier Lyapunov functions for the control of output-constrained nonlinear systems,''
\emph{Automatica},
vol. 45, no. 4, pp. 918-927, 2009.


\bibitem{Liu16}
Y. Liu, J. Li, S. Tong, and C. L. P. Chen,
``Neural network control-based adaptive learning design for nonlinear systems with full-state constraints,''
\emph{IEEE Trans. Neural Netw. Learn. Syst.},
vol. 27, no. 7, pp. 1562-1571, Jul. 2016.


\bibitem{Song17}
Y. Song, X. Huang, and Z. Jia,
``Dealing with the issues crucially related to the functionality and reliability of NN-associated control for nonlinear uncertain systems'',
\emph{IEEE Trans. Neural Netw. Learn. Syst.},
vol. 28, no. 11, pp. 2614-2625, Nov. 2017.


\bibitem{Ma15}
J. Ma, S. S. Ge, Z. Zheng, and D. Hu,
``Adaptive NN control of a class of nonlinear systems with asymmetric saturation actuators,''
\emph{IEEE Trans. Neural Netw. Learn. Syst.},
vol. 26, no. 7, pp. 1532-1538, Jul. 2015.

\bibitem{Tang16}
Z. Tang, S. S. Ge, K. P. Tee, and W. He,
``Robust adaptive neural tracking control for a class of perturbed uncertain nonlinear systems with state constraints,''
\emph{IEEE Trans. Syst., Man, Cybern.},
vol. 46, no. 12, pp. 1618-1629, Dec. 2016.

\bibitem{Song18}
Y. Song and S. Zhou,
``Neuroadaptive control with given performance specifications for MIMO strict-feedback systems under nonsmooth actuation and output constraints,''
\emph{IEEE Trans. Neural Netw. Learn. Syst.},
vol. 29, no. 9, pp. 4414-4425, Sept. 2018.











\bibitem{sadegh90}
N. Sadegh, R. Horowitz, W.-W. Kao, and M. Tomizuka, ``A unified
approach to the design of adaptive and repetitive controllers for robotic
manipulators," ASME J. Dyn. Syst., Meas., Control, vol. 112, no. 4,
pp. 618-629, Dec. 1990



\bibitem{yao09}
B. Yao, ``Desired compensation adaptive robust control," ASME J. Dyn. Syst. Meas. Control,
vol. 131, no. 6, pp. 1-7, Nov. 2009.

\bibitem{chenws08}
W. Chen and J. Li, ``Decentralized output-feedback neural control for systems with unknown
interconnections," IEEE Trans. Syst. Man Cybern. Part B Cybern., vol. 38, no. 1, pp. 258-266,
Feb. 2008.





\bibitem{sun18}
M. Sun, T Wu, L Chen, and G Zhang,
``Neural AILC for error tracking against arbitrary initial shifts,''
\emph{IEEE Trans. Neural Netw. Learn. Syst.},
vol. 29, no. 7, pp. 2705-2716, Jul. 2018.



\bibitem{sun23}
M. Sun and S. Zou, ``Adaptive learning control algorithms for infinite duration tracking,"
\emph{IEEE Trans. Neural Netw. Learn. Syst., }, vol. 34, no. 12,
pp. 10004-10017, Dec. 2023.



\bibitem{zou23}
S. Zou, M. Sun, and X. He,
``Integral Lyapunov function-based adaptive
learning control for nonstrict-feedback nonlinear systems'',
\emph{IEEE Trans.
Syst. Man Cybern.: Syst.},
vol. 53, no. 11, pp. 7152-7164, Nov. 2023.











\end{thebibliography}
}
\end{document}